\documentclass[11pt]{article}

\pdfoutput=1
\usepackage{amsmath}
\usepackage{amssymb}
\usepackage{graphicx}
\usepackage[english]{babel}
\usepackage[utf8x]{inputenc}
\usepackage[T1]{fontenc}
\usepackage{slashed}
\usepackage[usenames,dvipsnames]{xcolor}
\usepackage{fontawesome}
\usepackage{tikz}

\usepackage[colorinlistoftodos]{todonotes}
\usepackage[colorlinks=true,allcolors=blue]{hyperref}

\usepackage{bbold}
\usepackage{caption}
\usepackage{subcaption}
\usepackage{cancel}
\usepackage{framed}
\usepackage{comment}
\usepackage{mathtools}
\usepackage{chngcntr}
\usepackage{bigints}

\usepackage[square,numbers,sort&compress]{natbib}

\newcommand{\bra}[1]{\langle #1 |}
\newcommand{\ket}[1]{| #1 \rangle}

\newcommand{\nn}{\nonumber}
\newcommand{\be}{\begin{equation}}
\newcommand{\ee}{\end{equation}}
\newcommand{\bea}{\begin{eqnarray}}
\newcommand{\eea}{\end{eqnarray}}
\def\ltap{\ \raise.3ex\hbox{$<$\kern-.75em\lower1ex\hbox{$\sim$}}\ }
\def\gtap{\ \raise.3ex\hbox{$>$\kern-.75em\lower1ex\hbox{$\sim$}}\ }
\def\lsim{\ \raise.3ex\hbox{$<$\kern-.75em\lower1ex\hbox{$\sim$}}\ }
\def\gsim{\ \raise.3ex\hbox{$>$\kern-.75em\lower1ex\hbox{$\sim$}}\ }

\def\a{\alpha} \def\b{\beta}  \def\d{\delta}  
    
   \def\m{\mu} \def\n{\nu}
  \def\s{\sigma} \def\t{\tau}

\newcommand{\Mcal}{{\mathcal M}} \newcommand{\Ncal}{{\mathcal N}}
 \newcommand{\Ocal}{{\mathcal O}}

\newcommand{\bigstick}{\texttt{BIGSTICK}}

\newcommand{\nucleon}{\ensuremath{{\scriptscriptstyle N}}}
\newcommand{\nucleus}{\ensuremath{{\scriptscriptstyle\mathcal{N}}}}

\def\ltap{\ \raise.3ex\hbox{$<$\kern-.75em\lower1ex\hbox{$\sim$}}\ }
\def\gtap{\ \raise.3ex\hbox{$>$\kern-.75em\lower1ex\hbox{$\sim$}}\ }

\newcommand{\ra}{\rightarrow}

\usepackage{marginnote}

\begin{document}

\vspace*{1mm}

\noindent \makebox[9.5cm][l]{\small \hspace*{-.2cm} }{\small 
FERMILAB-PUB-23-781-T, IPMU23-0046}  \\  [-1mm]

\begin{center}
  {\Large \bf Earth-Catalyzed Detection of Magnetic Inelastic Dark Matter} \\[2mm]
  {\Large \bf with Photons in Large Underground Detectors}  \\

\vspace{10mm}

{Joshua Eby$^{1,2}$, Patrick J. Fox$^3$, and Graham D. Kribs$^4$}

\vspace*{5mm}

\noindent 
$^1$The Oskar Klein Centre, Department of Physics, Stockholm University,
    10691 Stockholm, Sweden \\[1mm]
$^2$Kavli IPMU (WPI), UTIAS, The University of Tokyo, Kashiwa, Chiba
    277-8583, Japan \\[1mm]
$^3$Particle Theory Department, Theory Division, Fermilab, Batavia, IL 60510, USA \\[1mm]
$^4$Institute for Fundamental Science and Department of Physics, \\
    University of Oregon, 
    Eugene, OR 97403 USA 
\end{center}

\vspace*{-5mm}

\begin{abstract}
  Inelastic dark matter with moderate splittings,
  $\mathcal{O}({\rm few} \; {\rm to} \; 150)$~keV,
  can upscatter to an excited state in the Earth, with the excited state
  subsequently decaying, leaving
  a distinctive monoenergetic photon signal in large
  underground detectors.  The photon signal can exhibit sidereal-daily modulation,
  providing excellent separation from backgrounds.
  Using a detailed numerical simulation, we examine this process as a search strategy for
  magnetic inelastic dark matter with the dark matter
  mass near the weak scale, 
  where the upscatter to the excited state and decay proceed through the
  same magnetic dipole transition operator.
  At lower inelastic splittings, the scattering is dominated by moderate 
  mass elements in the Earth with high spin, especially $^{27}$Al,
  while at larger splittings, $^{56}$Fe becomes the dominant target.  
  We show that the proposed large volume gaseous detector CYGNUS
  will have excellent sensitivity to this signal.  Xenon
  detectors also provide excellent sensitivity through the
  inelastic nuclear recoil signal, and if a future signal is seen,
  we show that the synergy among both types of detection can provide
  strong evidence for magnetic inelastic dark matter.  
  In the course we have calculated nuclear response functions for elements relevant for scattering in the Earth, which are publicly available on GitHub \href{https://github.com/joshaeby/IsotopeResponses}{\faGithub}.
\end{abstract}
\newpage
{
  \hypersetup{linkcolor=black}
  \tableofcontents
}


\section{Introduction}

Dark matter that scatters inelastically off nuclei is one of the
most promising weakly interacting massive particle (WIMP) candidates
\cite{Han:1997wn,Hall:1997ah,TuckerSmith:2001hy,TuckerSmith:2004jv,Finkbeiner:2007kk,Arina:2007tm,Chang:2008gd,Cui:2009xq,Fox:2010bu,Lin:2010sb,An:2011uq,Pospelov:2013nea,Dienes:2014via,Barello:2014uda,Bramante:2016rdh,Krall:2017xij}
that can substantially evade experimental constraints from
direct detection \cite{Bramante:2016rdh,Krall:2017xij,Song:2021yar}.
Over the past several years, research has continued along
several lines, including: astrophysical implications
such as heating up neutron stars \cite{Bell:2018pkk};
detection prospects at IceCube \cite{Catena:2018vzc};
detection of large inelastic splittings from
collisional de-excitation
\cite{Pospelov:2019vuf,Lehnert:2019tuw,Broerman:2020hfj};
a local enhancement of dark matter flux due to strongly-interacting
dark matter slowing down through multiple-scattering
off the atmosphere and/or Earth \cite{Pospelov:2019vuf};
detection of light dark matter that scatters inelastically
\cite{Tsai:2019buq} and through the Migdal effect
\cite{Bell:2021zkr};
new bounds from heavy element nuclear recoil or nuclear excitation
with very low background runs of detectors \cite{Song:2021yar};
cosmic-ray upscattered inelastic dark matter in the presence of a
light mediator \cite{Bell:2021xff}
as well as 
pseudo-Dirac dark matter through a
light mediator \cite{CarrilloGonzalez:2021lxm};
inelastic Dirac dark matter \cite{Filimonova:2022pkj};
the detection of luminous dark matter
when the dark matter is itself light enough to be amenable to
detection by sub-eV resolution solid-state detectors \cite{Bell:2022yxn};
production of gamma-rays for indirect detection~\cite{Berlin:2023qco}; formation of compact objects from inelastic dark matter cooling~\cite{Bramante:2023ddr}; direct searches at the LHC or beam-dump experiments~\cite{Dienes:2023uve,Asai:2023dzs,Jodlowski:2023ohn,Lu:2023cet}; capture in the Sun~\cite{Chauhan:2023zuf}; inelastic freeze-in production of dark matter~\cite{Heeba:2023bik}; searches using electron recoils in direct-detection experiments~\cite{Bell:2022dbf};
and various experimental searches for inelastic dark matter
including PandaX \cite{PandaX:2022djq},
Xenon1t \cite{XENON:2022avm},
and SENSEI \cite{Gu:2022vgb}.
Finally, several groups
\cite{Bell:2020bes,Bramante:2020zos,Baryakhtar:2020rwy,Bloch:2020uzh,Emken:2021vmf} also exploited inelastic dark matter in an attempt
to explain the now defunct Xenon1T electronic excess~\cite{XENON:2020rca}.


One of the most intriguing ideas for detecting inelastic dark matter
is through the upscatter of the excited state and its subsequent decay into
a photon.  This was first attempted in `luminous dark matter'
\cite{Feldstein:2010su} to explain the DAMA annual modulation,
and subsequently as dark matter detection through `two easy steps'
\cite{Pospelov:2013nea}.  The parameter space of these proposals,
however, has been ruled out by direct-detection
bounds on the nonobservation of the nuclear recoil from
upscattering of dark matter into its excited state \cite{Bramante:2016rdh}.
In 2019, we (with Harnik) \cite{Eby:2019mgs}
explored using the upscatter and subsequent decay to set bounds on
large inelastic splittings, of order hundreds of keV, where
the nuclear recoil sensitivity is severely limited by kinematics:
only dark matter with the very highest velocities were
capable of scattering off the moderately heavy elements
(xenon and iodine) used in the large detect-detection experiments.
We found upscatter off heavy elements in the Earth
(dominantly lead) provided a sufficient flux of
excited states such that Borexino was capable of setting bounds by using
the sidereal-daily modulation intrinsic to this signal \cite{Eby:2019mgs}.
The focus of the previous paper was dark matter inelastic scattering
off nuclei through a contact interaction, such as $Z$-exchange,
and subsequent decay to a photon, which arises for example in a
narrowly-split model of Higgsinos \cite{Fox:2014moa,Krall:2017xij}. 
Later work also explored the possibility that neutrinos, similarly, 
could upscatter and subsequently decay into photons
\cite{Plestid:2020vqf,Plestid:2020ssy}.

As discussed in \cite{Eby:2019mgs}, large gaseous detectors,
proposed for directional detection of WIMP dark matter, could
provide sensitivity also in a lower range of inelasticity where
direct detection experiments have increasingly better sensitivity
to the rate of nuclear recoil from upscattering.
The inelasticity range we found in \cite{Eby:2019mgs}
was from approximately $40$-$150$~keV, where it was possible that
the photon signal in the gaseous detectors could surpass the-then
bounds from xenon experiments.  
However, the model employed to analyze the sensitivity, 
namely upscatter through a contact interaction separate
from decay through a magnetic dipole transition operator, 
becomes somewhat baroque in this inelasticity range.
In particular, the magnetic dipole transition can easily
dominate over the contact interaction 
in the upscatter rate.
If this occurs, the sensitivity will be modified, since the decay
length is no longer independent of the upscatter rate in the Earth.

Thus, in this paper we focus on dark matter that can upscatter
off nuclei into an excited state exclusively through a magnetic dipole transition operator, that then decays via the same operator back
into dark matter and a photon.
In the theory where the magnetic dipole transition occurs 
as a result of a Dirac fermion is split into two Majorana states,
this is called Magnetic Inelastic Dark
Matter (MIDM) \cite{Kopp:2009qt,Chang:2010en}.
However, our results also apply to theories where the dark matter
and the excited state are a Dirac fermions with ``Dirac-to-Dirac''
magnetic dipole transition interaction.
These classes of models provide a concrete scenario in which as the strength of the magnetic dipole 
transition is varied, and therefore the resulting inelastic scattering rate off nuclei and the decay rate of the excited state are properly related.
The physics that we explore contains similarities to
our previous work \cite{Eby:2019mgs}: dark matter $\chi$ upscatters
to $\chi_2$ off an element $\Ncal$ in the Earth $\chi + \Ncal \ra \chi_2 + \Ncal$.
The excited state continues in the same general direction as the dark matter itself, for tens to thousands of kilometers, before
decaying inside a large underground detector.  This type of signal can exhibit 
a strong sidereal-daily modulation because when the detector is on the 
Cygnus-facing side of the Earth, there is far less ``target material''
available to upscatter. 

The signal we will consider therefore consists of single photons with an
energy equal to the mass splitting $\delta$ of inelastic dark matter, and
with a rate that can exhibit a strong modulation with a period of a sidereal
day.   The shape and phase of the modulation is predicted and depends on the
geographical location of the detector, allowing for excellent signal-to-background discrimination. 
Our focus in this paper will be on using a large proposed gaseous
detector, CYGNUS \cite{Vahsen:2020pzb}, which has been proposed to
reach $1000$~m$^3$ in volume.  The ostensible objective of a 
directional detector is to observe scattering off the nuclei that make up
the gas of the detector.  Our signal, however, is the spontaneous decay
of an excited state into dark matter and a photon, and so depends
only on the \emph{volume} of the detector, not the \emph{mass} of the gas
in the detector.  All we require is that there is a decent efficiency
to absorb the photon with few to hundreds of keV energy
in order to generate a scintillation signal.
Indeed, the ideal mode of operation of a large gaseous detector would be
to inject the lowest pressure gas that has decent efficiency to scintillate,
and choosing a gas that has exceptionally-low radioactive backgrounds
(and/or can be made exceptionally pure), such as helium.
In our analysis, we will consider the signal both in the presence of
radioactive backgrounds at a level that the CYGNUS collaboration reports can be straightforwardly achieved \cite{Vahsen:2020pzb}, as well as
a hypothetical mode of operation where the detector has been
fully optimized for the inelastic photon signal, i.e.,
background-free.  These will provide the realistic range of sensitivity
depending on the efforts to reduce backgrounds and the time in a
low-pressure, radiopure gas mode of operation.

We summarize the notation used in the paper in Table~\ref{tab:notation}.

\section{Dark Matter with a Magnetic Dipole Transition}
\label{sec:particlephysics}

Dark matter may be electrically neutral and yet possess
higher-dimensional interactions with the photon.
For fermionic dark matter, which is the focus of this paper,
the leading interaction at dimension-5 is a magnetic dipole
moment operator,
\begin{eqnarray} \label{eq:dipolemoment}
\frac{\tilde{\mu}_\Psi}{2} \bar{\Psi} \Sigma^{\mu\nu} \Psi F_{\mu\nu} \, ,
\end{eqnarray}
that we will show, as a byproduct of our analysis, is strongly
constrained from bounds on direct detection.
Here $\Psi$ is necessarily a Dirac fermion, $\Sigma^{\mu\nu}$
is a commutator of gamma matrices defined in Eq.~(\ref{eq:sigmamunu}),
and the coefficient of the magnetic dipole moment $\tilde{\mu}_\Psi$
has the dimensions of $[{\rm mass}]^{-1}$.

An alternative possibility is that a magnetic dipole moment for
dark matter is either absent or suppressed, and there is instead
a magnetic dipole transition to an excited state in the dark sector.
A well-known model that has a vanishing magnetic dipole moment but
allows for a magnetic dipole transition is
Magnetic Inelastic Dark Matter 
that consists of a 
Dirac fermion split into two
Majorana fermions \cite{Kopp:2009qt,Chang:2010en},
$\chi_2$ and $\chi$, with a small mass difference
$\delta \equiv m_{\chi_2} - m_\chi$.  We take $\delta > 0$,
and so $\chi$ is the dark matter while $\chi_2$ is an
excited state in the dark sector.  As we detail in
Appendix~\ref{app:emoperators}, the magnetic dipole 
transition between two Majorana fermions arises from 
the operator
\begin{eqnarray}
  i \frac{\tilde{\mu}_\chi}{2} \bar{\chi}_2\, \Sigma^{\m\n} \chi \,F_{\m\n}
  \, , 
\label{eq:dim5op}
\end{eqnarray}
where the magnetic dipole transition coefficient
$\tilde{\mu}_\chi$ has the units of $[{\rm mass}]^{-1}$. 

An alternative possibility is that there are two Dirac fermions
with magnetic dipole transition operators between them.  This arises in
the Standard Model, for example, where the neutral,
isospin-0 baryons $\Lambda_s^0$ and $\Sigma^0$
have both magnetic dipole moments and magnetic dipole transitions among them.
It is possible that dark sectors with dark baryons exist
where the magnetic dipole moments vanish,
and yet magnetic dipole transitions are present \cite{Asadi:2023toappear}.
The operators for the Dirac-to-Dirac magnetic dipole transitions have a
nearly identical form to the analogous Majorana-to-Majorana transition,
\begin{eqnarray}
  \frac{\tilde{\mu}_{12}}{2} \overline{\Psi}_{\mathrm{D}1} \Sigma^{\mu\nu} \Psi_{\mathrm{D}2}
  F_{\mu\nu}
  + \frac{\tilde{\mu}_{21}}{2} \overline{\Psi}_{\mathrm{D}2} \Sigma^{\mu\nu} \Psi_{\mathrm{D}1}
  F_{\mu\nu}
  \, ,
\label{eq:dim5opdirac}
\end{eqnarray}
as shown in Appendix~\ref{app:emoperators}. The
two transitions, $\Psi_{\mathrm{D}1} \rightarrow \overline{\Psi}_{\mathrm{D}2}$
and $\overline{\Psi}_{\mathrm{D}1} \rightarrow \Psi_{\mathrm{D}2}$
in Eq.~(\ref{eq:dim5opdirac}), are independent.  
If the magnetic dipole moment of dark matter vanishes ($\tilde{\mu}_{11} = 0$),
we can map the theory space of the Majorana-to-Majorana magnetic
dipole transition
precisely onto the Dirac-to-Dirac magnetic dipole transition.  
A simple way to understand the correspondence is to first
imagine that dark matter were entirely composed of $\Psi_{\mathrm{D}1}$
(and not $\bar{\Psi}_{\mathrm{D}1}$), which means the only magnetic dipole 
transition is $\Psi_{\mathrm{D}1} \ra \bar{\Psi}_{\mathrm{D}2}$.  The decay of
$\bar{\Psi}_{\mathrm{D}2} \ra \Psi_{\mathrm{D}1}$ proceeds through the same interaction.
As one shifts from a universe in which dark matter is made
entirely of $\Psi_{\mathrm{D}1}$ into, for example, a symmetric abundance
of $\Psi_{\mathrm{D}1}$ and $\bar{\Psi}_{\mathrm{D}1}$, these two ``species'' of dark matter
have two independent magnetic dipole transition operators that permit
the two species to scatter into
the excited states of $\bar{\Psi}_{\mathrm{D}2}$ and $\Psi_{\mathrm{D}2}$,
respectively.  If the dark sector preserves CP,
$\tilde{\mu}_{12} = \tilde{\mu}_{21}$, and so each ``half'' of the
dark matter (particles or antiparticles) has the same magnetic
dipole transition moment to an excited state as if there were only
particles, or equivalently, if the particle transition were
Majorana-to-Majorana.
Hence, our use of Eq.~(\ref{eq:dim5op}) for our calculations in the
paper applies to both a Majorana-to-Majorana model as well as
a Dirac-to-Dirac model, so long as the magnetic dipole moment of
dark matter vanishes.  

The magnetic dipole transition operator could arise from a perturbative
or nonperturbative ultraviolet (UV) completion.
A perturbative generation of the operator through an electroweak loop, integrating out some charged states of mass $m_{\pm}$, 
is expected to lead to $\tilde{\mu}_\chi \sim e g^2/(16\pi^2 m_{\pm})$, where $g$ is the electroweak coupling.  
Instead, if the dark matter is a composite of a new strongly-coupled sector,
the scale suppressing the operator is anticipated to be of order
the mass of the dark matter itself, and so there is no loop suppression; 
for example, this is the case for the magnetic dipole transition of
$\Sigma^0$ to $\Lambda_s^0$. 

Given that this interaction involves integrating out heavy
electrically charged particles, there is an upper bound on size
of the interaction by simply requiring that no new charged
particles are in conflict with the present LEP and LHC bounds.
The robust bound on charged particles comes from LEP II \cite{LEPchargebound1, LEPchargebound2},
where $m_{\pm} \gtrsim 100$~GeV\@. This bound is
essentially independent of lifetime.  There are stronger
bounds from the LHC in specific model-dependent scenarios.
We will generally consider $m_\chi = 1$~TeV for most of our
analyses, and so given that the dark matter is the lightest
particle in the dark sector, the charged states are necessarily
heavier, $m_{\pm} > m_\chi$.
We normalize the operator in Eq.~(\ref{eq:dim5op}) as 
\begin{eqnarray} \label{eq:mutilde}
\tilde{\mu}_\chi &=& \frac{g_M e}{4 m_\chi} \, ,
\end{eqnarray}
and so in practice the ratio $m_\chi/m_{\pm}$ is absorbed into our
definition of $g_M$. 
Given that we take $m_{\pm} > m_\chi = 1$~TeV,
this is fully sufficient to allow us to consider $g_M$ being either
perturbative, $g_M \sim 1/(16 \pi^2)$, or nonperturbative, $g_M \sim 1$.
Our sensitivity estimates,
and bounds from existing experiments, will be presented as
bounds on $g_M$.

The magnetic dipole transition operator in Eq.~(\ref{eq:dim5op})
allows for several interesting processes including upscattering off nuclei
and excited state decay.  (Scattering off electrons will be negligible
for the dark matter masses we are interested in.)  Our focus will be on 
upscatter of dark matter off nuclei in the Earth, followed by excited state decay
into a monoenergetic photon, discussed in the following sections.
There is also the possibility that the excited state could downscatter off
nuclei, although this process is suppressed compared with spontaneous decay.\footnote{
Downscattering through the dipole operator Eq.~(\ref{eq:dim5op}) is not kinematically suppressed, and, ignoring form factor effects (that only suppress the rate), the cross section is $\sigma_{\mathrm{ds}}\sim (Ze)^2 \tilde{\mu}_\chi^2$.  In an environment with nuclear density $n$ the typical distance traveled before a down-scatter event occurs is $\ell_{\mathrm{ds}}\sim (n\sigma_{\mathrm{ds}})^{-1}$.  Comparing this to the typical distance traveled before a decay Eq.~(\ref{eq:decaylength}) we find that for $\delta \gtrsim 1$~keV and above range the downscattering process is negligible.} 

\subsection{Excited state decay}

\begin{figure}[t]
\centering
\includegraphics[scale=0.5]{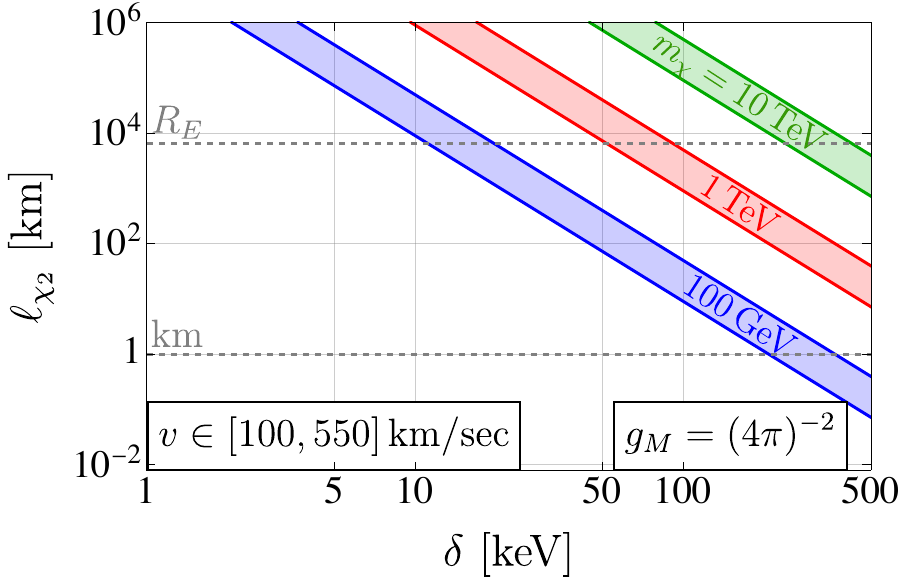}\quad
\includegraphics[scale=0.5]{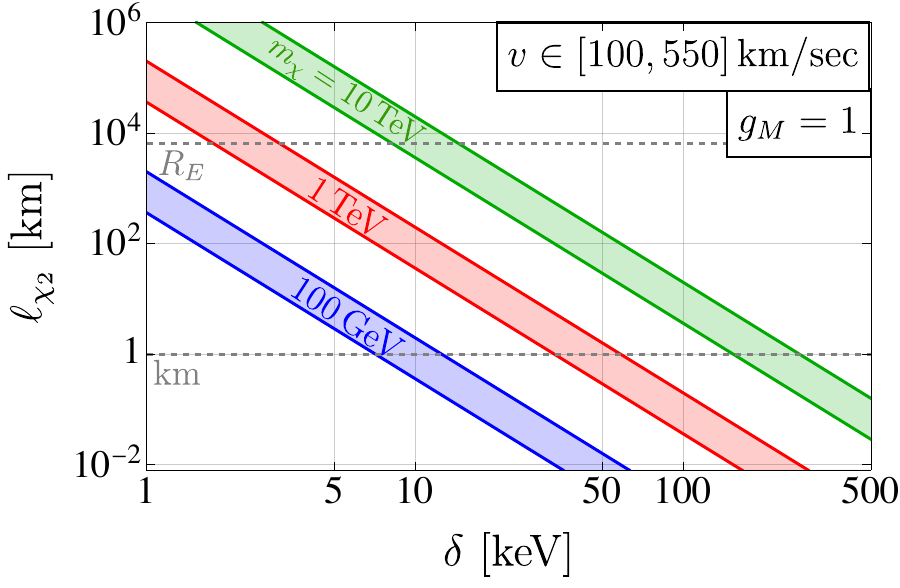}
\caption{Decay length of excited state, Eq.~\eqref{eq:decaylength}, as a function of $\d$,
  for $m_{\chi} = 100$~GeV, $1$~TeV, and $10$~TeV,
  for weakly-coupled (left) and strongly-coupled (right) theories;
  the extent of the shaded regions denote variation of
  $v$ in the range $100-550$~km/sec.}
\label{fig:decaylength}
\end{figure}

The heavier dark matter state can decay through emission of a photon,
$\chi_2\rightarrow \chi \gamma$.  The width of the excited state is
\begin{eqnarray}\label{eq:excitedwidth}
  \Gamma_{\chi_2} \;\equiv\; \frac{1}{\tau_2} &=& \frac{\tilde{\mu}_\chi^2}{\pi} \delta^3
                      \qquad (\delta \ll m_\chi) \, , 
\end{eqnarray}
where this decay rate applies to both the Majorana-to-Majorana 
as well as the Dirac-to-Dirac magnetic dipole transition-mediated decay. 
The lifetime of the excited state is $\tau_2$, and so 
when it moves with speed $v$ in the Earth frame, the 
typical distance $\ell_{\chi_2}$ it will travel is 
\begin{equation}\label{eq:decaylength}
  \ell_{\chi_2} \;=\; v \, \tau_2 \; \simeq \;
 \left( 1 \; {\rm km} \right) \times 
 \left(\frac{0.1}{g_M}\right)^2\left(\frac{m_\chi}{\rm TeV}\right)^2
 \left(\frac{250\,{\rm keV}}{\d}\right)^3
 \left(\frac{v}{450\,{\rm km/sec}}\right) \, .
\end{equation}
The decay length is illustrated
in Figure \ref{fig:decaylength} for both perturbative (left)
and nonperturbative (right) values of $g_M$. Importantly, splittings $\delta$ 
below $\mathcal{O}({\rm few})$~keV will be below the threshold of foreseeable
experimental searches, and above $\mathcal{O}(500)$~keV is also problematic due
to the initial upscatter becoming kinematically disallowed
(see \cite{Eby:2019mgs}). Distances of order km and the Earth's radius,
$R_E$, are illustrated
for reference.  When $\ell_{\chi_2} \gtrsim R_E$, the number of viable targets
will drop significantly as most decays take place outside the volume
of the Earth. On the other hand, when $\ell_{\chi_2} \lesssim 1$~km, the overburden becomes
nearly isotropic in the vicinity of the detector, suppressing the
modulating nature of our signal. Therefore the region between the
gray dashed lines, where ${\rm km}\lesssim \ell_{\chi_2}\lesssim R_E$, is optimal. 

\subsection{Up-scattering}

The magnetic dipole transition operator leads to dark matter scattering
off Standard Model (SM) particles\@.  A detailed and thorough
analysis of this process is given
in \cite{DelNobile:2021wmp}.
For direct detection, the scattering of dark matter
off a nucleon $N = (p, n)$ occurs
through the insertion of two photon currents,
\begin{equation}
  \langle N'|e J_{em}^\mu | N \rangle
  \frac{g_{\mu\nu}}{q^2}
  \langle \chi_2 | J_{\rm MDT}^\nu | \chi \rangle \, ,
\end{equation}
here $N'$ denotes the same nucleus as $N$ but with a different momentum.  The first term is just what one would expect for the electromagnetic
coupling of a nucleon, 
\begin{eqnarray}
\label{eq:nucleon-photon-coupling}
  \langle N' |e J_{em}^\mu | N \rangle
  &=& e \overline{u}_{N'} \left[ F_1(q^2) \gamma^\mu
      + F_2(q^2) \frac{i\sigma^{\mu\nu}q_\nu}{2m_N} \right] u_N \, \\
  &=& e \overline{u}_{N'} \left[ F_1(q^2) \frac{p^\mu}{2 m_N} 
      + \Big( F_1(q^2) + F_2(q^2) \Big)
        \frac{i\sigma^{\mu\nu}q_\nu}{2m_N} \right] u_N \, 
\end{eqnarray}
where $p^\mu \equiv p_N^\mu + p_{N'}^\mu$,
$q^\mu \equiv p_{N'}^\mu - p_{N}^\mu$, 
and $F_1(q^2)$ and $F_2(q^2)$ are the conventional electromagnetic
form factors of the nucleon with $F_1(0) = Q_N$ and
$F_2(0) = \kappa_N$.  The full magnetic dipole moment of the
nucleon is $g_N \hat{\mu}_N s_N$, with $\hat{\mu}_N \equiv e/(2 m_N)$
(the nucleon magneton) and $s_N = 1/2$, and so we have
\begin{equation}
  g_N = 2 (Q_N + \kappa_N) \simeq \left\{
    \begin{array}{ll}
      + 5.59 & \quad {\rm proton} \\
      - 3.83 & \quad {\rm neutron}         
    \end{array}
  \right. \, .
\end{equation}
The second term arises from the magnetic dipole transition (MDT)
\begin{eqnarray}
  \langle \chi_2 | J_{\rm MDT}^\mu | \chi \rangle
  &=& \tilde{\mu}_\chi \overline{u}_{\chi_2} i \Sigma^{\mu\nu} q_\nu u_\chi \, .
\label{eq:DM-photon-coupling}
\end{eqnarray}

Since the speed distribution for dark matter in our halo has a cutoff
at an escape velocity $v_{\rm esc}/c \ll 1$, we can take the
non-relativistic expansion of the spinors $u_i$ and $\bar{u}_i$ in
Eqs.~(\ref{eq:nucleon-photon-coupling}),(\ref{eq:DM-photon-coupling})
and then match to the set of inelastic scattering operators presented
in Ref.~\cite{Barello:2014uda}. 
The only operators generated
are $\mathcal{O}_{1,4,5,6}$ which are defined as
\begin{eqnarray}
\label{eq:operators}
  \mathcal{O}_1  &=& \mathbb{1} \\
  \mathcal{O}_4  &=& \vec{S}_\chi \cdot \vec{S}_N \\
  \mathcal{O}_5  &=& i \vec{S}_\chi \cdot \left(\frac{\vec{q}}{m_N}
                     \times \vec{v}^\perp \right) \\
  \mathcal{O}_6  &=& \left(\vec{S}_\chi \cdot \frac{\vec{q}}{m_N}\right)
                     \left(\vec{S}_N \cdot \frac{\vec{q}}{m_N}\right)~.
\end{eqnarray}
Here, $\vec{v}^\perp \equiv \vec{v} + \vec{q}/(2 \mu_N) $,
$\vec{S}_\chi$ is the dark matter spin vector,
and $\vec{S}_N$ is the nucleon spin vector.
As emphasized in \cite{Fitzpatrick:2012ix}, these constitute a
complete set of Galilean and Hermitian
invariants.\footnote{Strictly, $i \vec{q}$
  is Hermitian \cite{Fitzpatrick:2012ix}.} 
The operators $\mathcal{O}_1$ and $\mathcal{O}_4$ are the typical 
spin-independent and spin-dependent dark-matter--nucleon couplings,
respectively.  The operator $\mathcal{O}_{5}$ is the coupling of
dark matter spin with $\vec{q} \times \vec{v}^\perp$,
which can be thought of as an analogue of the spin-orbit coupling
$\vec{S}_\chi \cdot \vec{L}_N$, treating the orbital angular momentum
of the nucleon as $\vec{L}_N = \vec{r} \times \vec{p}$ with
$\vec{p} = m_N \vec{v}$ and $\vec{r} = \vec{q}/m_N^2$. 
The operator $\mathcal{O}_{6}$, analogous to a tensor force,
will turn out not to contribute to the scattering in our case, as we will see below. 
The amplitude for dark matter scattering
off nucleons through the magnetic dipole transition operator is
\begin{equation}\label{eq:DMnucleonME}
  \mathcal{M}_{\chi,N} = c_1^N \mathcal{O}_1 + c_4^N \mathcal{O}_4
                         + c_5^N \mathcal{O}_5 +c_6^N \mathcal{O}_6 \, ,
\end{equation}
with the $c_i^N$ given by
\begin{eqnarray}
c_1^N &=& \frac{Q_N e \tilde{\mu}_\chi}{2m_\chi} ~, \label{eq:c1} \\
c_4^N &=& \frac{g_N e \tilde{\mu}_\chi}{m_N} ~,\label{eq:c4} \\ 
c_5^N &=& -\frac{2 Q_N e \tilde{\mu}_\chi m_N}{q^2} ~,\label{eq:c5} \\
c_6^N &=& -\frac{g_N e \tilde{\mu}_\chi m_N}{q^2} \label{eq:c6} ~ .
\end{eqnarray}
The operator coefficients are given in terms of the nucleon charge
$Q_N = (1, 0)$ and magnetic dipole moments
$g_N \simeq (5.59, -3.83)$ for the (proton, neutron)
respectively.  It will also be useful to write the $c_i$
in terms of isospin quantities:
\begin{eqnarray}
  c_i^0 &\equiv& \frac{c_i^p + c_i^n}{2} ~,\label{eq:ci0} \\
  c_i^1 &\equiv& \frac{c_i^p - c_i^n}{2} \label{eq:ci1} ~ .
\end{eqnarray}

Going from the DM-nucleon coupling Eq.~(\ref{eq:DMnucleonME}) to the DM-nucleus coupling requires
a model for how nucleons behave inside nuclei.
We follow \cite{Anand:2013yka} and assume that the DM-nuclear interaction
can be described by the individual DM-nucleon interactions summed over
all nucleons in the nucleus.  The insertion of the nucleon operators into
the nucleus means that the operators $\mathcal{O}_i$ must be evaluated
for bound nucleons and the DM as a plane wave.  These operators can
be multipole-decomposed and the final spin-averaged, non-relativistic matrix element squared can be
factorized as 
\begin{eqnarray} 
  \frac{1}{(2\,j_\chi + 1)(2\,j_\nucleus + 1)}
  \sum_{\rm spins} \left| \Mcal_{\rm NR} \right|^2 
  &=& \frac{4\pi}{2\,j_\nucleus+1}\sum_k \sum_{\substack{\t=\,0,1 \\ \t'=\,0,1}}
      R_k^{\t\t'}
      \left(v_\nucleus^{\perp\,2},\frac{q^2}{m_\nucleon^2},\d,c_i\right)\,
      W_{k}^{\t\t'}(y) \, ,
\label{eq:MNRsqu}
\end{eqnarray}
where $j_\nucleus$ is the spin of the nucleus,
$v_\nucleus^{\perp\,2} = v_\chi^2 - v_{{\rm min}\,T}^2$ with
$v_{{\rm min}\,T}^2 = \left(\frac{q^{2}}{2\mu} + \delta\right)^2$
being the minimum dark matter speed necessary to scatter off the
nucleus with recoil energy $E_R= q^2/(2m_\nucleus)$, and $\mu$ is the dark-matter--nucleus reduced mass. 
In general the sum over $k$ depends upon eight DM response functions,
$R_k$, and nuclear response functions $W_k$.  The DM response functions
depend upon the details of the particle physics through the $c_i$
and the nuclear response functions are determined by expectation values
of the allowed six nuclear operators in the nucleus.\footnote{We assume
  P and CP are good symmetries of the nucleus and these six correspond
  to vector charge, vector transverse magnetic, axial transverse electric,
  axial longitudinal, vector transverse electric, and vector longitudinal
  operators.}
  
The DM response functions for inelastic dark matter with general couplings
were presented in \cite{Barello:2014uda}.  For the case of the MIDM
with the $c_i$ as given in Eqs.~(\ref{eq:c1})-(\ref{eq:c6}) only four
DM response functions are non-zero.\footnote{For general non-zero
  $c_{1,4,5,6}$ one would also expect $R_{\Sigma''}$ to also contribute,
  but for the $c_i$ generated by a magnetic dipole transition
  operator, $R_{\Sigma''}=0$.}  These correspond to $k=M, \Sigma', \Delta, \Delta\Sigma'$ and are
\begin{eqnarray}
  R_M^{\tau\tau'}
  &=& c_1^\tau c_1^{\tau'}
      + \frac{|\vec{q}|^2}{4m_n^2} c_5^\tau c_5^{\tau'}
      \left( v_\nucleus^2 - v_{{\rm min}\,\nucleus}^2 \right) ~,\\
R^{\tau\tau'} _{\Sigma'}
  &=&  \frac{c_4^\tau c_4^{\tau'}}{16}~, \\
R^{\tau\tau'} _\Delta
  &=& \frac{c_5^\tau c_5^{\tau'}|\vec{q}|^2}{4m_\nucleon^2} ~,\\ 
R^{\tau\tau'} _{\Delta\Sigma'}
  &=& \frac{c_5^\tau c_4^{\tau'}}{4} ~.
\end{eqnarray}
Here the $\tau, \tau'$ are isospin labels and $c_i^\tau$ are
given in Eqs.~(\ref{eq:ci0}),(\ref{eq:ci1}).

\begin{table}[t]
\renewcommand{\arraystretch}{1.3}
\begin{center}
\begin{tabular}{c|l} \hline\hline
  $\chi$ & fermionic dark matter candidate \\
  $\chi_2$ & excited (neutral) dark sector state \\
  $\delta$ & mass splitting, $\delta\equiv m_{\chi_2}-m_\chi$ \\
  $\tilde{\mu}_\chi$ & coefficient of magnetic dipole transition operator, 
             Eq.~(\ref{eq:dim5op}) \\
  $\vec{S}_\chi$ & dark matter spin vector 
                   \\ \hline
  $N$, $m_N$, $Q_N$
    & nucleon [$N=(p,n)$], nucleon mass, nucleon charge \\
  $\mu_N$ & DM / nucleon reduced mass, $\mu_N \equiv m_\chi m_N/(m_\chi + m_N)$ \\
  $g_N$ & nucleon magnetic dipole moment \\ 
  $\vec{S}_N$ & nucleon spin vector \\ 
  $c^N_i$ & model-dependent nucleon coefficients \\ \hline 
  $\mathcal{N}$, $M_{\mathcal{N}}$, $Q_{\mathcal{N}}$, $A$
    & nucleus, nucleus mass, nucleus charge, nucleus mass number \\
  $\mu$ & DM / nucleus reduced mass, $\mu \equiv m_\chi m_{\Ncal}/(m_\chi + m_{\Ncal})$ \\
  $\mu_2$ & $\chi_2$ / nucleus reduced mass, $\mu_2 \equiv m_{\chi_2} m_{\Ncal}/(m_{\chi_2} + m_{\Ncal})$ \\
  $j_{\mathcal{N}}$, $T$, $M_T$
    & nucleus spin, total nucleus isopin, $z$-component of nucleus isospin \\ 
  $R^{\tau\tau'}_k$, $W^{\tau\tau'}_k$
  & DM response function, nuclear response function \\
    \hline\hline
\end{tabular}
\end{center}
\caption{Notation used in the paper.}
\label{tab:notation}
\end{table}

The determination of the corresponding nuclear response functions requires knowledge of the nuclear wave functions,  since the $W_k$ are expectation values of operators in nuclear states,
\begin{eqnarray}
W_{\Ocal^{A}\Ocal^{B}}^{\t\t'}
  &=& \sum_{J=0}^\infty \bra{j_\nucleus} | \Ocal^{A}_{J;\t} | \ket{j_\nucleus}
      \bra{j_\nucleus} | \Ocal^{B}_{J;\t'} | \ket{j_\nucleus} \, .
\end{eqnarray}
The technical details of this procedure can be found in Appendix \ref{app:OBME}, which follows \cite{Fitzpatrick:2012ix, Anand:2013yka, DelNobile:2021wmp}.  The outcome is that the nuclear response functions are simple polynomials (times an exponential) of the exchanged momentum $q$.  For instance, the only non-zero response function for $^{24}$Mg is $W_M^{00}$, which is
\begin{equation}
W_M^{00}= e^{-2 y} \left(0.123 \left(9.63-7.49 y+ y^2\right)^2\right) \, ,    
\end{equation}
with $y\approx (4.6\,q/\mathrm{keV})^2$.

Once the DM and nuclear response functions have been determined, the differential scattering cross section is given by 
\begin{eqnarray}
\label{eq:dsigmadcostheta}
 \frac{d\s^{Z,A}_{\rm MDT}}{d\cos\theta^{\mathrm{cm}}}(v^2,q^2) &=& \frac{\mu^2}{4\pi}\sqrt{1-\frac{2\delta}{\mu v^2}}\frac{1}{(2\,j_\chi + 1)(2\,j_\nucleus + 1)}\sum_{\rm spins} \left| \Mcal_{\rm NR} \right|^2 \, ,
\end{eqnarray}
where $\cos\theta^{\mathrm{cm}}$ is the cosine of the center-of-mass scattering angle. 
One can compute the total scattering
rate of MIDM with $N_T$ scattering targets
per unit mass as
\begin{eqnarray} \label{eq:dRdER}
  \frac{dR^{Z,A}}{dE_R}
  &=& \frac{\rho_\chi N_T}{m_\chi}
      \int_{v_{\rm min}}^{v_{\rm max}}
      d^3v_{\rm MB} \,v \,f_{\rm gal}(v_{\rm MB}) \frac{d\sigma^{Z,A}_{\rm MDT}}{dE_R}(v^2,q^2) \, ,
\label{eq:dRdER}
\end{eqnarray}
where $v_{\rm min}$ and $v_{\rm max}$ bound the kinematically-allowed
speeds (see e.g.~\cite{Bramante:2016rdh} for details).
For simplicity we will assume the velocity distribution, in galactic coordinates, is given by a Maxwell-Boltzmann distribution $f_{\rm gal}(v_{\rm MB})$ with escape speed of $v_{\rm esc}=550$ km/s and velocity dispersion of $220$ km/s. To transform from the galactic frame velocity $\vec{v}_{\rm MB}$ to the Earth frame velocity $\vec{v}$, we follow the series of Galilean boosts and rotations outlined in \cite{Eby:2019mgs}.  We take the local DM density to be $\rho_\chi = 0.3$ GeV/cm$^3$.
The cross section $d\sigma^{Z,A}_{\rm MDT}/dE_R$ is proportional to
Eq.~\eqref{eq:dsigmadcostheta}, 
and is easily derived from the relation (see e.g. Appendix D of~\cite{DelNobile:2021wmp} for a full derivation),
\begin{equation}
    E_R = \frac{\mu^2 v^2}{m_\nucleus}\left(1 - \cos\theta^{\mathrm{cm}}\sqrt{1-\frac{2\delta}{\mu v^2}}\right) 
            - \frac{\mu \delta}{m_\nucleus} \,.    
\end{equation}
For a given element the rate is 
calculable using the product of response functions in
Eq.~\eqref{eq:MNRsqu}.

\begin{figure}[t]
\centering
\includegraphics[scale=0.5]{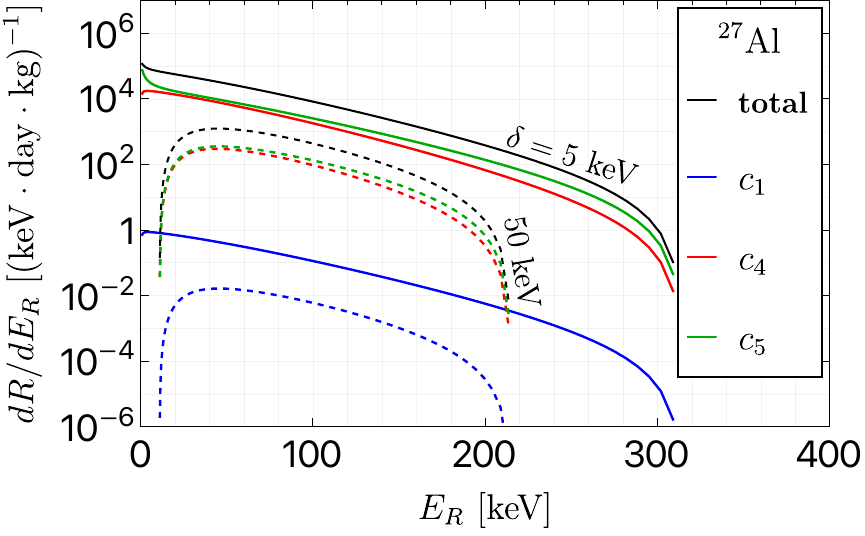} \quad
\includegraphics[scale=0.5]{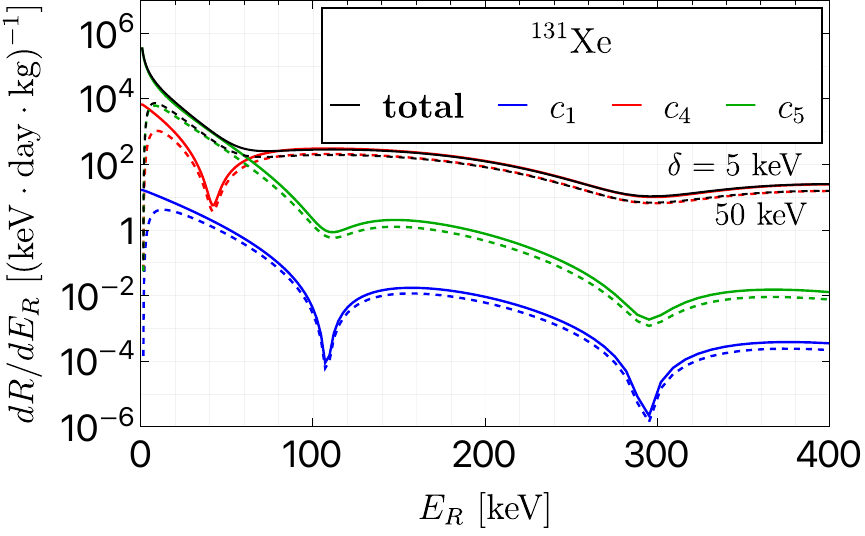}
\caption{Comparison of $c_i$ contributions to two isotopes 
  of relevance to us: $^{27}{\rm Al}$ (spin 5/2) and $^{131}{\rm Xe}$
  (spin 3/2).}
\label{fig:alxerates}
\end{figure}

As an example, we show in Figure~\ref{fig:alxerates} the
differential scattering rates for two isotopes that will be
of significant relevance for us: $^{27}{\rm Al}$ and $^{131}{\rm Xe}$. 
We show the total differential scattering rate (that includes interference among contributions), as well as
the individual contributions from the different coefficients (which are individually squared to get the sub-component rates)  
$c_1$, $c_4$, and $c_5$, for two specific values of the
inelasticity $\delta = 5,50$~keV\@.  
Because both elements have non-zero spin
($5/2$ and $3/2$ respectively), $c_4$ is non-zero.
One of the critical observations in Figure~\ref{fig:alxerates}
is that the scattering rates of both of these elements receive a large contribution from $c_5$.  For $^{27}{\rm Al}$, $c_5$ provides the dominant contribution for most recoil energies, while $c_4$ (and the interference between $c_4$ and $c_5$) also make a significant contribution to the scattering rate.
For $^{131}{\rm Xe}$, $c_5$ dominates for $E_R \lesssim 50$~keV, whereas the response is dominated by the spin-dependent contribution $c_4$ for $E_R \gtrsim 50$~keV\@.
Notice also that in both cases the spin-independent contribution,  $c_1$, is negligible.  The recoil energy-dependent (and $\delta$-dependent) nuclear responses are one of the characteristics of theories that scatter through a magnetic dipole transition. 


\section{Signal Rate Calculation}

The calculation of the expected event rate from dark matter upscattering off a nucleus in the Earth and subsequently decaying into a photon in a large volume underground detector was presented in \cite{Eby:2019mgs}. Here we review the inputs to the calculation as they pertain to dark matter scattering through the magnetic dipole transition operator. 

The event rate involves integrating over all possible scattering sites in the Earth and requires knowing the elemental composition of the Earth.   Unlike our previous calculation \cite{Eby:2019mgs},  
the nontrivial dependence of the magnetic dipole transition cross section on several nuclear response functions means that we need to know the specific \emph{isotopic} abundances of materials in the Earth, rather than the more straightforward \emph{atomic} abundances.  We denote the number density for a particular element with atomic number $Z$ and mass number $A$ as $n^{Z,A}(r)$, as a function of the radius $r$.  For the radial dependence of the abundances, we employ a simplified three-layer model of the Earth consisting of a core, mantle, and crust.  Thus, our number densities are piecewise-constant functions of radius.  In particular, we take the Earth to be a sphere of radius $6371$~km with the core and mantle having outer radii of $3483$~km and $6341$~km, respectively.  

We extract the isotopic abundances for each element from an online database hosted
by Brookhaven National Laboratory \cite{nuclearwallet} and remove any isotope whose lifetime is shorter than the age of the Earth. Where possible, we cross-checked the results with~\cite{STONE200575}.
We then convolve these isotopic fractions, denoted $f$, with the known terrestrial abundances for the elements in the core \cite{core}, mantle \cite{mantle}, and crust \cite{crust} to arrive at the isotopic number density in each of the three regions of the Earth.  We have collected these and other important quantities for stable isotopes on Earth in an ancillary file available on GitHub~\cite{Eby_IsotopeResponses}.

\begin{figure}[t]
\centering
\includegraphics[scale=0.5]{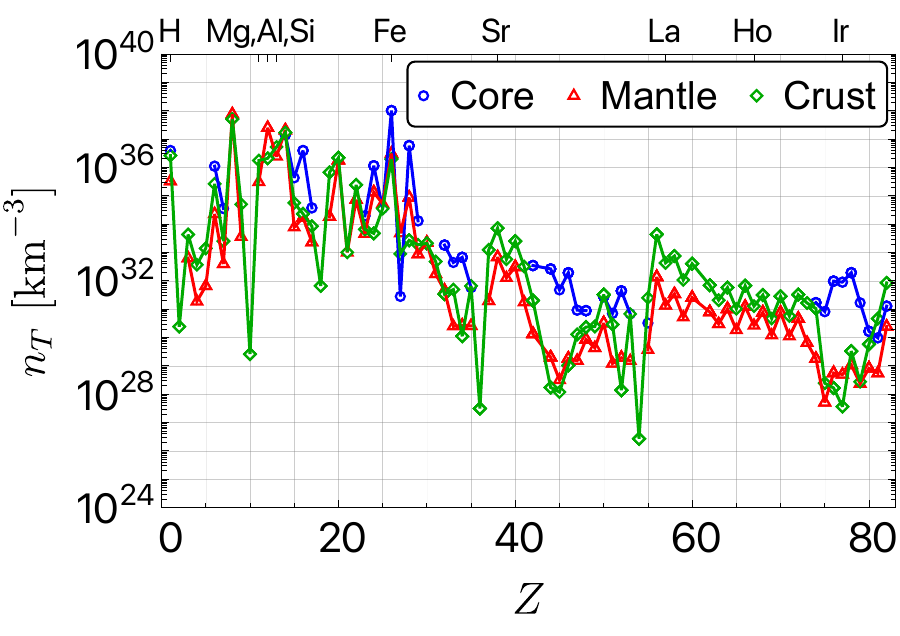} \quad
\includegraphics[scale=0.5]{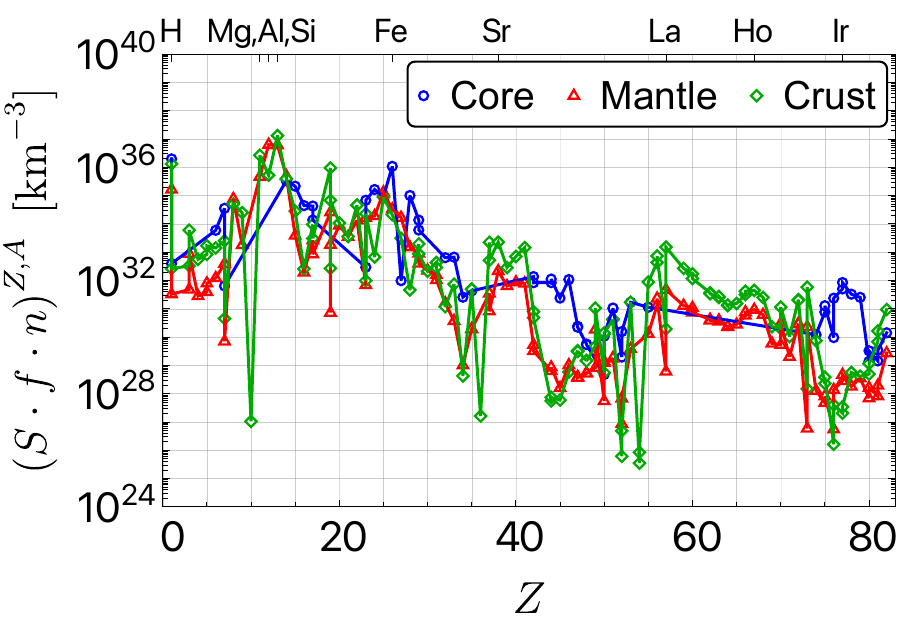}
\caption{Number density (left) and spin-weighted isotopic fraction
  times number density for elements with non-zero densities on Earth (right)
  for each atomic number $Z$ \cite{core,mantle,crust}.}
\label{fig:targets}
\end{figure}

As we have seen, the magnetic dipole transition scattering cross section is a recoil energy-dependent combination of several nuclear response functions. In our numerical calculations the full dependence is taken into account.  However, as a qualitative guide for which elements in the Earth will provide the dominant contributions to the rate, we present in Figure~\ref{fig:targets} the number density $n_T$ for each element
(aggregated over all isotopes) (left panel), and the spin-weighted quantity, $S\times f\times n^{Z,A}$, for each isotope (right panel).  These are approximate measures of the density or number of spin-independent and spin-dependent targets, respectively.  From these plots we expect that the rate will be dominated by contributions from ${}^{24}$Mg,${}^{25}$Mg,${}^{27}$Al,${}^{29}$Si, and ${}^{56}$Fe.  Note that $^{57}$Fe is another element that gives a non-zero contribution to the rate, especially for larger inelastic splittings.  However, we found a high level of sensitivity of the nuclear response functions to choices about the details of the nuclear physics model for this particular nucleus, and so we did not include this in our calculations, making our bounds conservative. 
We present, in Table~\ref{tab:isotopes}, the parameters of the five elements that dominate the signal rate and are used in our calculations.

\begin{table}[t]
\renewcommand*{\arraystretch}{1.2}
 \begin{center}
 \begin{tabular}{c|  c   c  c  c  c c}
  \hline\hline
   Element & Spin, Parity  & Isotopic & $\tilde{\mu}_{\nucleus}$ & $n_{\rm core}$ & $n_{\rm mantle}$ & $n_{\rm crust}$ \\[-0.2em]
           &               & fraction $(f)$ & $[e/(2 m_p)]$ & $[{\rm km}^{-3}]$ & $[{\rm km}^{-3}]$ & $[{\rm km}^{-3}]$ \\ \hline
$^{24}_{12}\mathrm{Mg}$ & $0,+$ & $0.79$ & $0$ & $0$ & $2.49 \times 10^{37}$ & $2.08 \times 10^{36}$ \\ 
$^{25}_{12}\mathrm{Mg}$ & $5/2,+$ & $0.10$ & $-0.855$ & $0$ & $2.49 \times 10^{37}$ & $2.08 \times 10^{36}$ \\ 
$^{27}_{13}\mathrm{Al}$ & $5/2,+$ & $1.00$ & $+3.642$ & $0$ & $2.40 \times 10^{36}$ & $5.24 \times 10^{36}$ \\ 
$^{29}_{14}\mathrm{Si}$ & $1/2,+$ & $0.05$ & $-0.555$ & $1.41 \times 10^{37}$ & $2.06 \times 10^{37}$ & $1.68 \times 10^{37}$ \\ 
$^{56}_{26}\mathrm{Fe}$ & $0,+$ & $0.92$ & $0$ & $1.01 \times 10^{38}$ & $3.07 \times 10^{36}$ & $1.97 \times 10^{36}$ \\ 
 \hline\hline
\end{tabular}
\caption{For each of the elements used to calculate the rate, we show the
  spin/parity, isotopic fraction (by number density), magnetic dipole moment
  (in units of the nuclear magneton, $\frac{e}{2m_p}$), and the number
  densities per cubic kilometer in the core, mantle, and crust.
  Additional information for more elements is provided in the ancillary
  materials.  This data was compiled from Refs~\cite{core,mantle,crust,nuclearwallet}, which report uncertainties not greater than $10\%$ for the isotopes relevant to this work.}
\label{tab:isotopes}
\end{center}
\end{table}

In the centre-of-mass frame the scattering of the DM is isotropic, see Eq.~(\ref{eq:dsigmadcostheta}), and the kinematics is straightforward.
The outgoing speed of the excited DM state after a nuclear scatter is
\begin{equation}
  v_{\mathrm{out}}^{\mathrm{cm}}
  = \left[ \frac{\mu_{2}}{m_{\chi_2}^2}
           \left(\mu  v^2-2\delta\right) \right]^{1/2}~,
\end{equation}
with $\mu$ the reduced mass of the DM and the nuclear target, and similarly for $\mu_{2}$ with $m_\chi$ replaced by $m_{\chi_2}$.  In the Earth's frame (equivalently the lab frame), the scattering angle is related to that in the center-of-mass frame by
\begin{equation}
v_{\mathrm{out}}^{\mathrm{lab}} \cos\theta^{\mathrm{lab}} = v_{\mathrm{out}}^{\mathrm{cm}}\cos\theta^{\mathrm{cm}} + \frac{\mu v}{m_\nucleus}~.
\end{equation}

The direction that the DM must scatter to reach the detector is determined by the relative position of the scatter site and the detector.  This one direction in the lab frame could arise from two possible scattering angles in the center-of-mass frame, both of which result in $\chi_2$ arriving at the detector.  However, these correspond to two \emph{different} speeds for $\chi_2$ in the Earth frame, following the upscatter process $\chi + \Ncal \rightarrow \chi_2 + \Ncal$.  These two different speeds are
\begin{equation}
\label{eq:voutlab}
v_{\mathrm{out},\pm}^{\mathrm{lab}} = \frac{\mu v}{m_\nucleus} \cos\theta^{\mathrm{lab}} \left(1\pm 
\left[1-\frac{1-\left(\frac{m_\nucleus v_{\mathrm{out}}^{\mathrm{cm}}}{\mu  v}\right)^2}{\cos^2\theta^{\mathrm{lab}}}\right]^{1/2}\right)~.
\end{equation}
For heavy dark matter, $m_\chi > m_\nucleus$, the scattering is preferentially forward and the lab scattering angle $\theta^{\mathrm{lab}}$ is kinematically limited, with the maximum\footnote{When $m_\nucleus v_{\mathrm{out}}^{\mathrm{cm}} > \mu  v$, as typically occurs for dark matter lighter than the nucleus, all scattering angles are available and $\theta_{\mathrm{max}}^{\mathrm{lab}} = \pi$.} possible scattering angle denoted $\theta_{\mathrm{max}}^{\mathrm{lab}}$.  The cosine of this maximum scattering angle is  
\begin{equation}
  \cos^2\theta_{\mathrm{max}}^{\mathrm{lab}}
  = 1 - \left(\frac{m_\nucleus v_{\mathrm{out}}^{\mathrm{cm}}}{\mu v}\right)^2
  			    ~.
\label{eq:costhetamaxiDM}
\end{equation}
Focusing on the case of heavy dark matter, we use Eq.~(\ref{eq:costhetamaxiDM}) to express the outgoing speeds in the lab frame in a convenient form
\be
v_{\mathrm{out},\pm}^{\mathrm{lab}} = \frac{\mu v}{m_\nucleus}
\cos\theta^{\mathrm{lab}}
	  \left[1 \pm \sqrt{1 - \frac{\cos^2{\theta^{\mathrm{lab}}_{\mathrm{max}}}}{\cos^2{\theta^{\mathrm{lab}}}}}\,\right]~.
\ee
In order for $\chi_2$ to arrive at the detector, the direction from the scattering site (at position $\vec{r}_s$) to the detector (at position $\vec{r}_D$) must lie within a cone subtended by opening angle $\theta_{\mathrm{max}}^{\mathrm{lab}}$, and the fraction of the cone that the detector covers is given by $[R_D/(|\vec{r}_s-\vec{r}_D| \theta_{\mathrm{max}}^{\mathrm{lab}})]^2$, where we assume the detector is spherical with radius $R_D$.  Note that the amount of available scattering material grows as $|\vec{r}_s-\vec{r}_D|^2$ while the likelihood of scattering towards the detector scales as $|\vec{r}_s-\vec{r}_D|^{-2}$. Thus, all scatter sites within a decay length of the detector are of (approximately) equal importance. However, at distances larger than the typical decay length $\ell_{\chi_2}$, more distant scatter sites become less important.

\begin{figure}[t]
\centering
\includegraphics[scale=0.7]{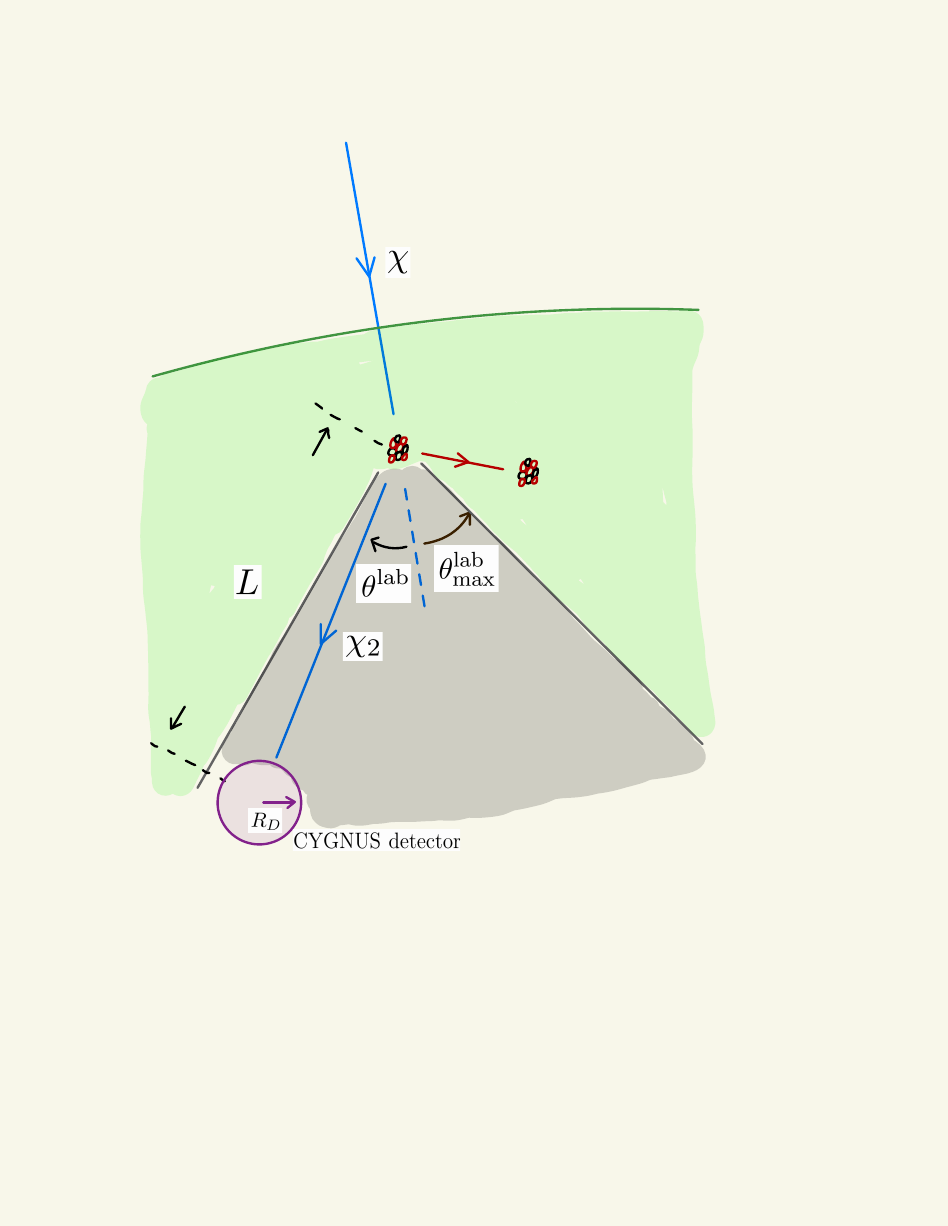} \caption{Illusration of the dark matter upscatter and lab frame kinematical quantities.  The green region represents the Earth, while the gray region represents a 2-d projection of the cone subtended by opening angle $\theta_{\rm max}^{\rm lab}$.  In this figure, only one (of two) choices of $v_{{\rm out},\pm}^{\rm lab}$ was taken.}
\label{fig:labframeillustration}
\end{figure}

The two different lab-frame speeds $v_{\mathrm{out},\pm}^{\mathrm{lab}}$ lead to two different decay lengths and consequently have different probabilities to decay in the detector.
The probability that an excited state with lifetime $\tau_2$ moving at speed $v_\mathrm{out}^{\mathrm{lab}}$ will travel a distance $L = \left|\vec{r}_s - \vec{r}_D \right|$ and decay inside the detector is
 \begin{equation}
 P(v_\mathrm{out}^{\mathrm{lab}}, L,\tau_2) = 2\,\sinh\Big(\frac{R_D}{2\,v_\mathrm{out}^{\mathrm{lab}}\tau_2}\Big)\exp\left(-\frac{L}{v_\mathrm{out}^{\mathrm{lab}}\tau_2}\right)~.
 \end{equation}
We illustrate the lab-frame kinematics for one choice of the outgoing speed in Figure~\ref{fig:labframeillustration}.

Furthermore, for the two possible center-of-mass scattering angles there are two different values of the exchanged momentum, $q_\pm$, resulting in two different scattering cross sections. When calculating the total rate we sum over both of these possibilities.  Integrating over all scattering angles in the lab frame requires introducing a Jacobian to relate the center-of-mass frame, where the scattering is isotropic, to the lab frame,
\begin{equation}
   J_\pm(v) = \frac{d\cos\theta^{\mathrm{cm}}}{d\cos\theta^{\mathrm{lab}}} \nonumber \\
      = 2\,z\,\cos\theta^{\mathrm{lab}} \pm 
	      \frac{1 - z^2 + 2\,z^2\,\cos^2\theta^{\mathrm{lab}}}
		    {\sqrt{1 - z^2 + z^2\,\cos^2\theta^{\mathrm{lab}}}},
\end{equation}
with $z= \mu v/(m_\nucleus\, v_{\mathrm{out}}^{\mathrm{cm}})$.

Combining all of the effects discussed above we arrive at the final expression for the signal rate for scattering off one particular isotope in the Earth, 
\begin{eqnarray}
  \label{eq:totalrateMIDM}
 \Gamma^{Z,A} = &\displaystyle{
        \sum_{\pm}\bigintss d^3r_s\,d^3v_{\mathrm{MB}}\,
 		\bigg\{ n^{Z,A}(r_s)\,\frac{\rho_\chi}{m_{\chi}} 
		\left[\frac{R_D}{\left|\vec{r}_s - \vec{r}_D\right|\,\theta_{\mathrm{max}}^{\mathrm{lab}}}\right]^2\,
		P(v_{\mathrm{out},\pm}^{\mathrm{lab}}, L, \tau_2)
            } \nonumber \\
        & \displaystyle{
        \times v\, f_{\mathrm{gal}}(v_{\mathrm{MB}})\,
        \frac{d\sigma^{Z,A}_{\rm MDT}(v^2,q_\pm^2)}{d\cos\theta^{\mathrm{cm}}}\left|J_{\pm}(v)\right|\,
        		 \bigg\}
            }~.
\end{eqnarray}
The total rate will be a sum over all of the relevant isotopes in the Earth.  In what follows we limit ourselves to 
a subset of all elements and focus on those which will give the dominant contribution to the rate, which are those summarized in Table~\ref{tab:isotopes}.  We are always interested in the limit where the detector size is smaller than the decay length $R_D\ll \ell_{\chi_2}$.  Hence the rate scales as the volume of the detector and is independent of its geometry.

\subsection{Rates at a typical detector}

The rate of scattering is sensitive to the path length ($L$) in Eq.~(\ref{eq:totalrateMIDM}) that the dark matter must travel to reach the detector.  This means that the signal rate at underground detectors is (sidereal) time-dependent~\cite{Eby:2019mgs}, with a maximum rate at the time that the Cygnus constellation is furthest below the horizon. This is because the DM flux appears in Earth frame to be coming from the direction of the constellation Cygnus, and the rate is maximized when there is the greatest number of scattering targets within a typical decay length \emph{in front of} the underground detector. The significant time-dependence of the rate -- sidereal-daily modulation -- depends upon the latitude of the laboratory, the distribution of elements within the Earth (Table~\ref{tab:isotopes}), and the dark matter parameters.  In our analysis we consider two locations for detectors: Gran Sasso, Italy at 42.6$^\circ$ North and the Stawell Underground Physics Laboratory (SUPL), Australia  at 37.1$^\circ$ South.  The Gran Sasso underground lab is at a depth of approximately $1.4\, \mathrm{km}$; it contained the (now decommissioned) Borexino experiment and is a possible future location for a detector belonging to the worldwide  CYGNUS directional dark matter experiment \cite{CYGNUSww}.  SUPL, whose depth is around $1\,\mathrm{km}$, is the current home of the SABRE-South experiment \cite{SABRETDR,SABRE:2018lfp} as well as another possible home for a CYGNUS detector.  The Cygnus constellation is below the horizon for about $2.5$ hours for Gran Sasso and $18.5$ hours at SUPL.

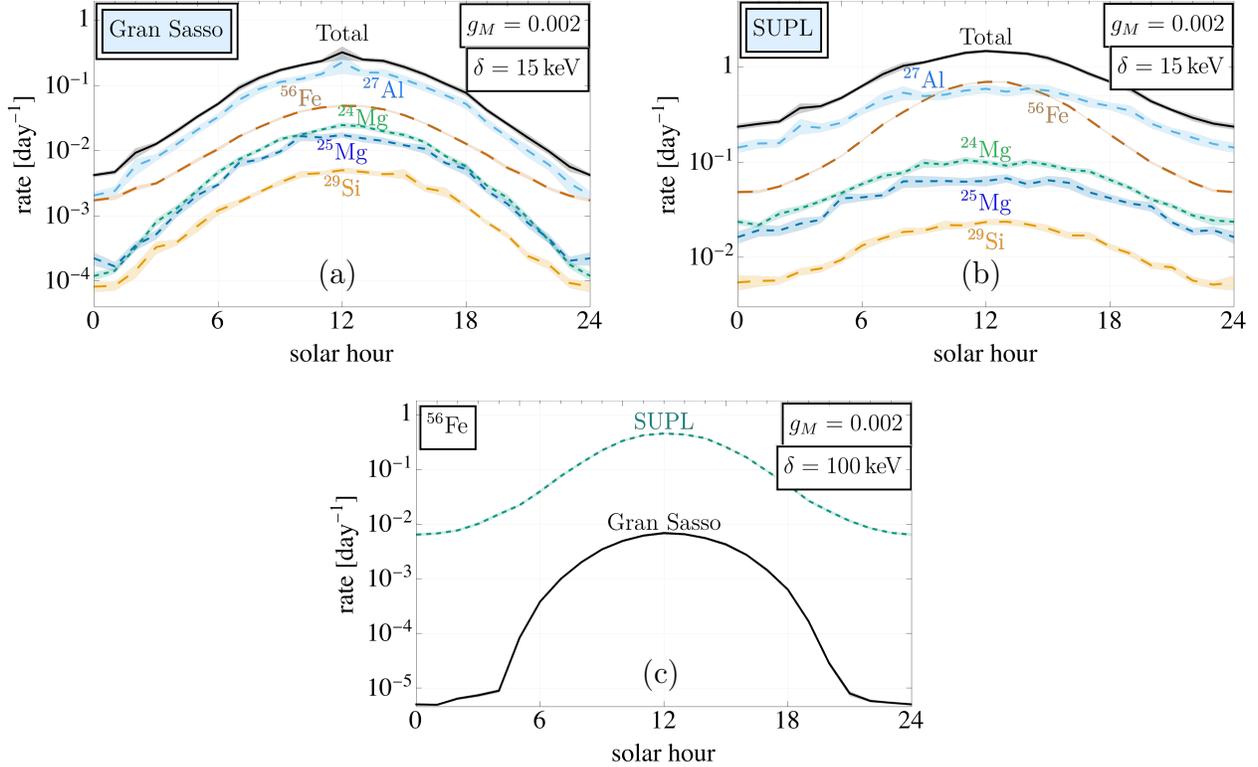
\begin{figure}[t]
\centering
\begin{tikzpicture}
\draw (0,0) node[inner sep=0] {\includegraphics[width=0.48\linewidth]{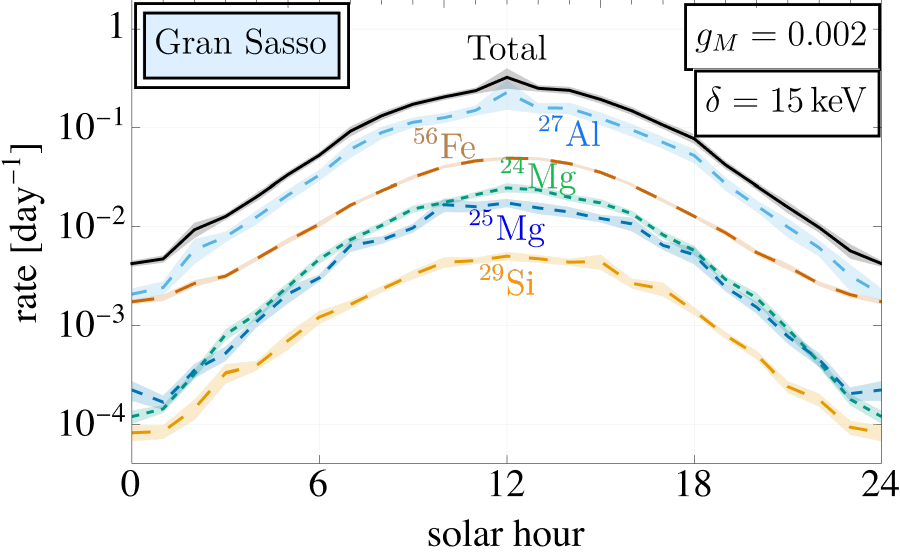}
};
\draw (0.43,-1.2) node {\large{(a)}}; 
\end{tikzpicture}
\hfill
\begin{tikzpicture}
  \draw (0,0) node[inner sep=0]
{\includegraphics[width=0.48\linewidth]{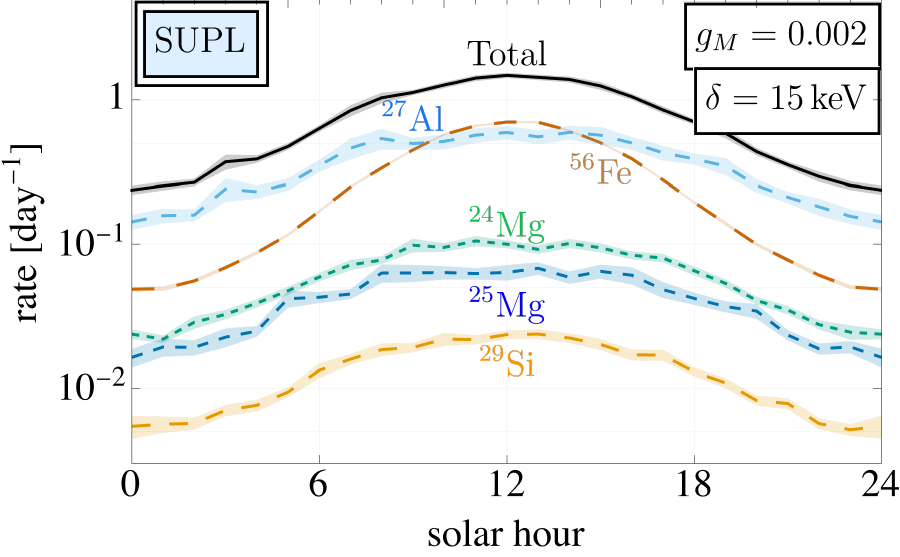}}; 
\draw (0.43,-1.2) node {\large{(b)}}; 
\end{tikzpicture} \\
\vspace{4mm}
\begin{tikzpicture}
  \draw (0,0) node[inner sep=0]
  {\includegraphics[width=0.48\linewidth]{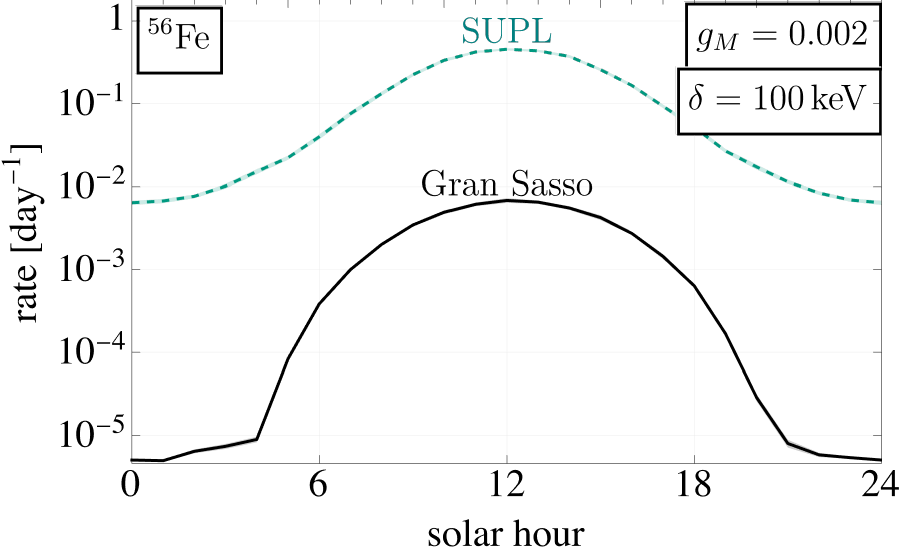}};
\draw (0.45,-1.2) node {\large{(c)}}; 
\end{tikzpicture}
\caption{Daily modulation rates, using Eq.~\eqref{eq:totalrateMIDM}, at an underground detector placed in 
  Gran Sasso (a) or SUPL (b) for dark matter of mass $1\,\mathrm{TeV}$
  and splitting $\delta=15\,\mathrm{keV}$.  In (c) the splitting is 
  $\delta=100\,\mathrm{keV}$ and the only scattering takes place off iron.  Solar hour was normalized so that the peak of the scattering rates occurs in the middle of the plot; the actual solar hour corresponding to the peak scattering rate will vary from solar hour 0 to 24 over the course of a year.}
\label{fig:modulationplot}
\end{figure}

Using as inputs the information about the distribution of isotopes in the Earth, Table~\ref{tab:isotopes}, along with the scattering cross sections and lifetime of the excited state discussed in Section~\ref{sec:particlephysics}, we can calculate the signal rate using Eq.~(\ref{eq:totalrateMIDM}).  For both Gran Sasso and SUPL we assume the underground detector has a volume of  $1000\,\mathrm{m}^3$, as proposed for a future CYGNUS detector \cite{Vahsen:2020pzb}.  As an illustration of the expected event rate for a de-excitation photon to be produced in a detector at each location, we show the rate in Figure~\ref{fig:modulationplot} as a function of time throughout a single day, which we choose to be January $1$.  The phase of the daily modulation in the figure is arbitrary, with the peak chosen to line up with dark matter ``noon'' (which is not the same as local terrestrial noon).  The solar time corresponding to the peak rate varies throughout the solar year, due to the difference between a sidereal day (approximately $23$ hours and $56$ minutes long) and a solar day.

In Figure~\ref{fig:modulationplot}(a) we show the rate at Gran Sasso for DM of mass $m_\chi = 1\,\mathrm{TeV}$, inelastic splitting $\delta = 15\,\mathrm{keV}$, and magnetic dipole transition coefficient $g_M=2\times 10^{-3}$.  For these DM parameters, the typical decay length of the excited state is considerably larger than the size of the Earth; see Figure~\ref{fig:decaylength}. This means the event rate grows linearly with the path length of the dark matter through the Earth.
We show the rate for various elements but it is dominated at this latitude by scattering off elements in the crust, in particular aluminum.  In Figure~\ref{fig:modulationplot}(b) we show the signal expected at SUPL for the same DM model.  Over much of day the rate is again dominated by scattering off aluminum but the rate is an order of magnitude larger than at Gran Sasso due to the increased path length that DM takes through the Earth, since the Cygnus constellation is in the northern
hemisphere while SUPL is in the southern hemisphere.  At the time of highest rate, when Cygnus is lowest in the sky, some of the DM that de-excites in the detector has passed through the core of the Earth and the rate from iron briefly dominates (since there is very little aluminum in the core of the Earth, see Table~\ref{tab:isotopes}).  Note that in comparing the rate shape at Gran Sasso to that at SUPL, the peak is broader at SUPL because Cygnus is below the horizon for a longer fraction of the day at SUPL in the southern hemisphere.  In Figure~\ref{fig:modulationplot}(c) we show the rate for DM with the same mass and magnetic dipole transition coupling but with an inelastic splitting of $\delta=100\,\mathrm{keV}$.  With these parameters the decay length is shorter but is still comparable to the radius of the Earth; see Figure~\ref{fig:decaylength}.  For such a large splitting, the only possible scattering (among the elements we consider) is off iron, and again the rate at SUPL is larger than at Gran Sasso due to the increased overburden.

\begin{figure}[t]
\centering
\includegraphics[scale=0.39]{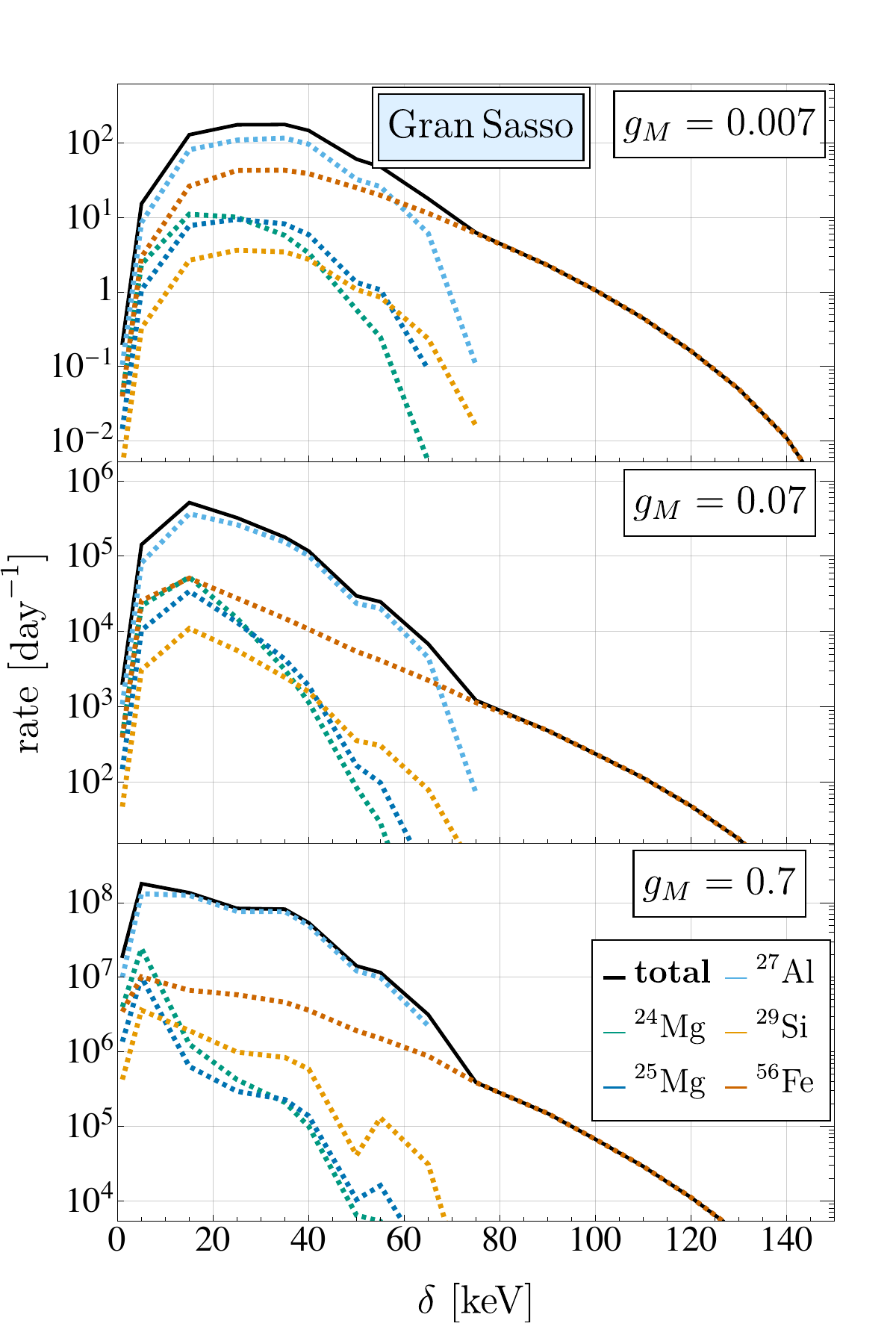} \quad
\includegraphics[scale=0.39]{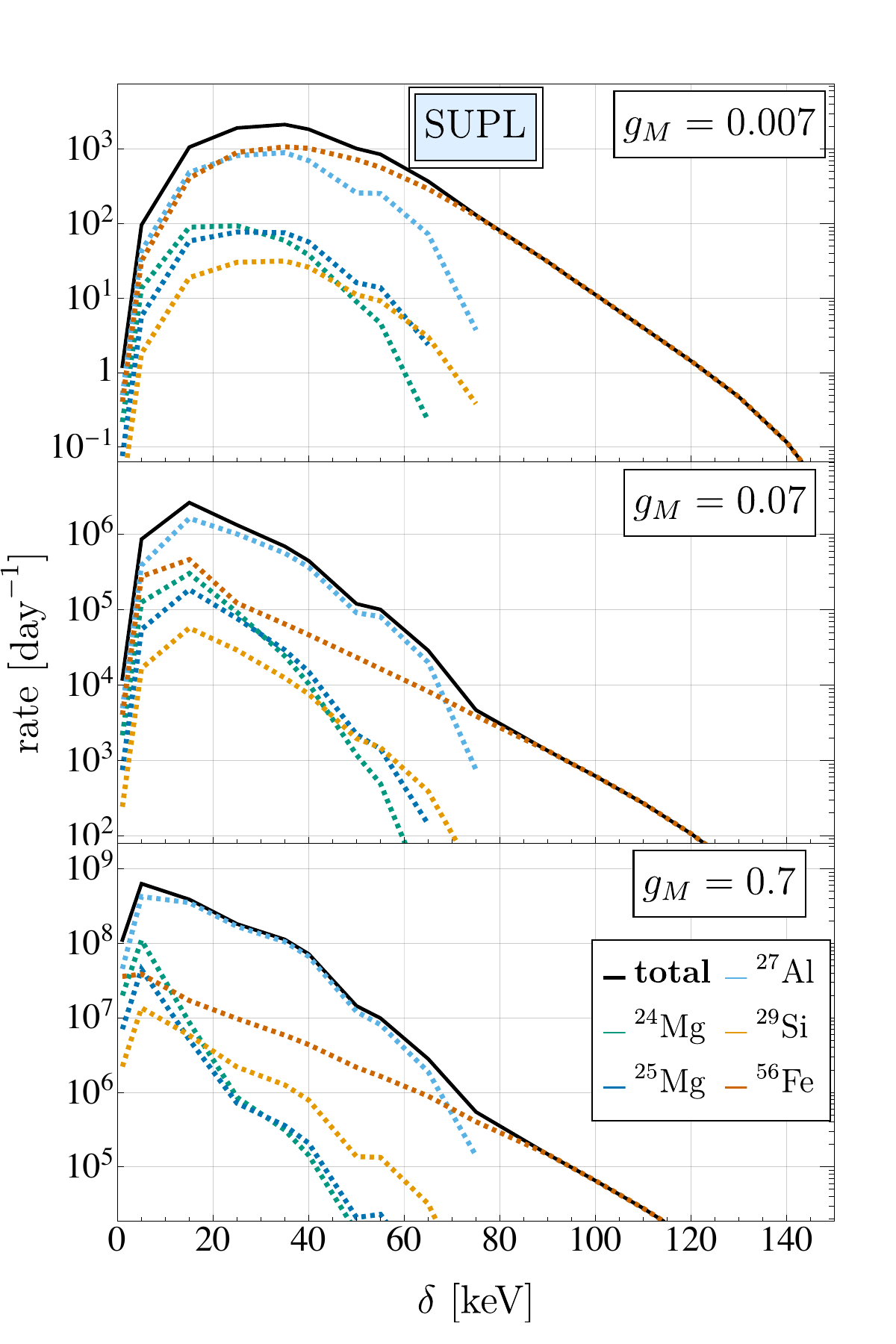}
\caption{Total daily rate, using Eq.~\eqref{eq:totalrateMIDM},  
    for scattering off
  different elements at Gran Sasso (left) and SUPL (right) as a function of mass splitting $\delta$
  and size of the magnetic dipole transition coefficient $g_M$.}
\label{fig:elementplot}
\end{figure}

In Figure~\ref{fig:elementplot} we present a series of plots that show how the total daily rate, integrated over a sidereal day, varies as both mass splitting $\delta$ and size of magnetic dipole transition coefficient $g_M$ are changed.  There is a intricate interplay between the scattering cross sections, which scales as $g_M^2$ [Eq.~(\ref{eq:dsigmadcostheta})], the width of the excited state, which scales $g_M^2 \delta^3$ [Eq.~(\ref{eq:excitedwidth})], and the location in the Earth of the elements where the upscatter transition $\chi + \Ncal \rightarrow \chi_2 + \Ncal$ occurs.  At large mass splitting and large coupling, where the distance traveled by $\chi_2$ is $\ltap \mathcal{O}(1)\,\mathrm{km}$, there is little difference between the rates expected at SUPL and Gran Sasso.  At large mass splitting (when the scattering is mostly forward, off iron, and DM needs a high speed to upscatter) and small coupling the rate at SUPL is greater than that at Gran Sasso due to the increased overburden.  Furthermore, the rate at SUPL scales as $g_M^2$ whereas at Gran Sasso it scales as $\sim g_M^{2.4}$, since there is a smaller fraction of overburden that is optimal for $\chi_2$ decaying within Gran Sasso.  
At lower $\delta$ where scattering can occur off more elements the signal scales as a higher power of $g_M$.  The rapid decrease in signal at low mass splitting is due to the $\chi_2$ lifetime becoming longer than the diameter of Earth.

\section{Sensitivity to MIDM}

We have seen so far that the total signal rate of photons from MIDM decay can be quite large,
with many events occurring per day and a substantial sidereal-daily
modulation effect, over the range
$\mathcal{O}({\rm few})\,{\rm keV} \lesssim \delta \lesssim 150\,{\rm keV}$
and for $g_M\ll 1$. Now we move to estimate the sensitivity to
this signal in future experiments, as well as existing constraints on
these parameters.  In Section~\ref{ssec:CYGNUS}, we estimate the sensitivity
of a future CYGNUS (or similar) gaseous detector to this monoenergetic photon signal.  In Section~\ref{ssec:LZ}, we recast present
direct detection constraints on the same parameter space.

\subsection{Sensitivity of CYGNUS} \label{ssec:CYGNUS}

\begin{figure}[t]
\centering
\includegraphics[scale=0.5]{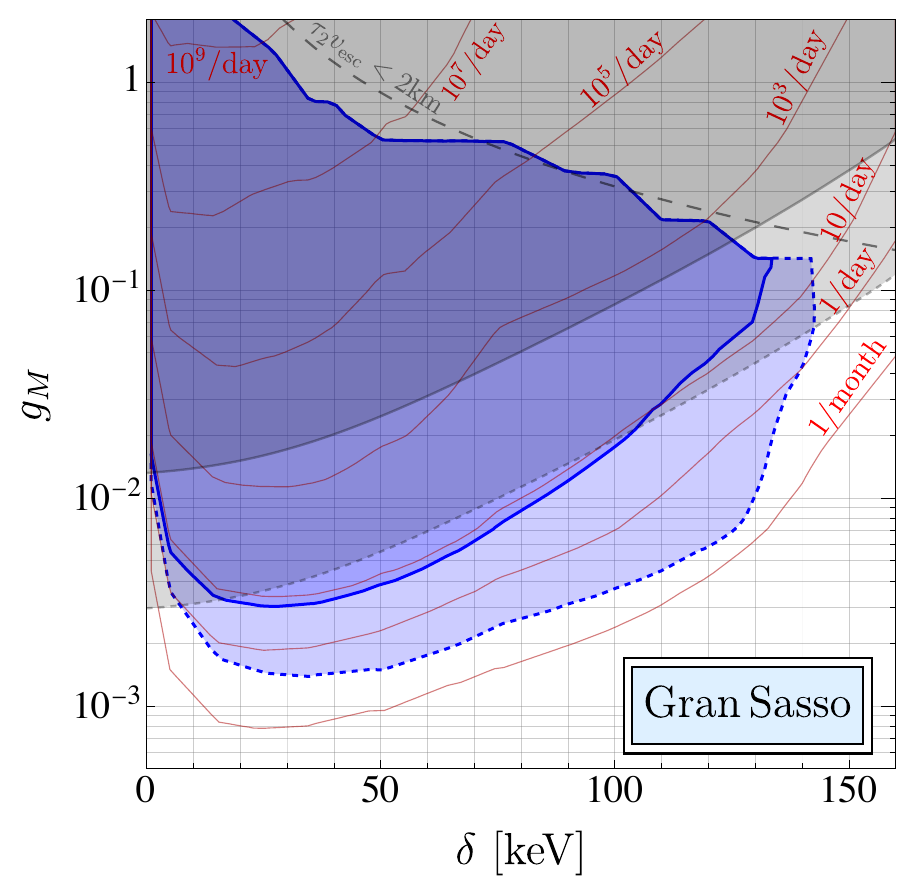} \quad
\includegraphics[scale=0.5]{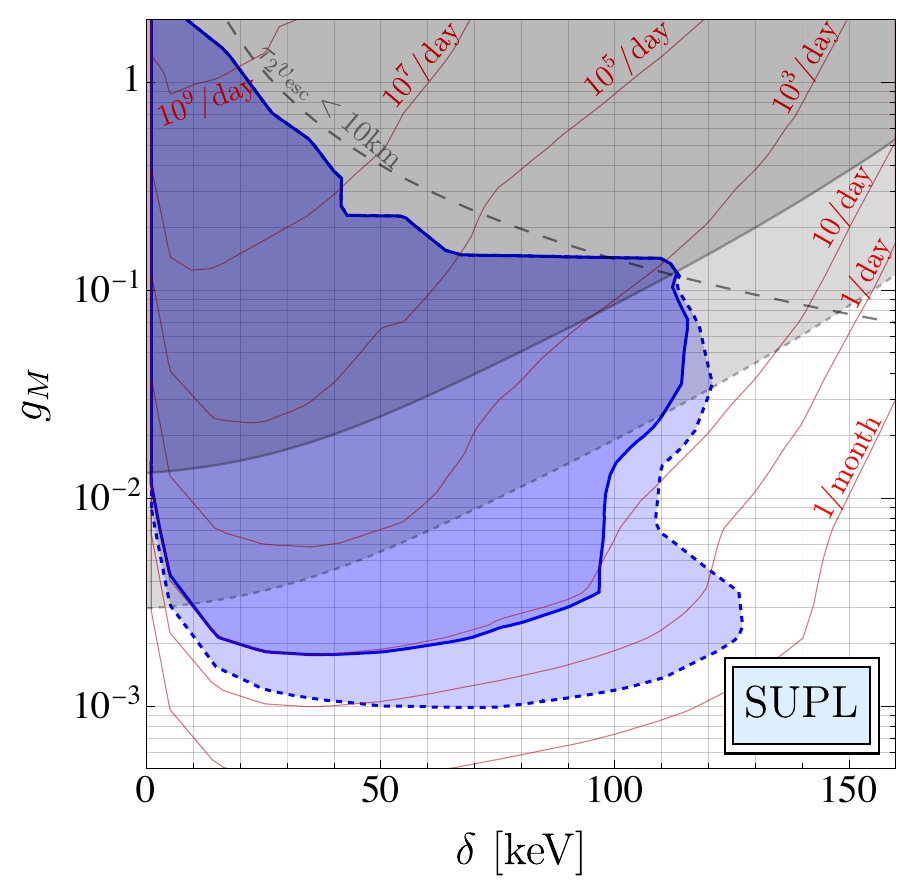}
\caption{Sensitivity estimate for the modulating MIDM signal in a
  $1000\,{\rm m}^3$ CYGNUS detector located at Gran Sasso (left) and
  SUPL (right), using $t_{\rm int}=6\,{\rm year}$ with the background
  rate projection of~\cite{Vahsen:2020pzb} (dark-blue shaded region),
  and $t_{\rm int}={\rm year}$ with zero background (light-blue shaded region).
  The light gray (dark gray) region is the excluded
  by LZ~\cite{LZ:2022lsv} by requiring the number of signal events $N_{\rm sig} \lesssim 16.6$ ($356$). The red contours give the total rate in the detector
  volume, and the dashed gray lines correspond to decay lengths roughly
  of order the overburden at each location; see text for details.}
\label{fig:sensitivity}
\end{figure}

We calculate the sensitivity of a search for the photon signal 
arising from MIDM at the
position of Gran Sasso as well as SUPL, which are two proposed locations
under consideration for the eventual CYGNUS detector \cite{CYGNUSww}. We assume
a fiducial volume of $1000$~m$^3$ and a total experimental integration time
of $t_{\rm int}=6$~years, as projected in~\cite{Vahsen:2020pzb}.
The results are given in Figure~\ref{fig:sensitivity} and explained below.

The total signal rate (i.e.~averaging modulation out) is the integral
of Eq.~\eqref{eq:totalrateMIDM} over the total integration time and is shown by the red contours in Figure~\ref{fig:sensitivity}. In practice, we simply
sum over the rate in Eq.~\eqref{eq:totalrateMIDM} in 1-hour bins to obtain
the total rate/day; see e.g.\ Figure~\ref{fig:elementplot}.  Multiplying by the expected number of days of
planned CYGNUS observation ($N_{\rm days} = 6\cdot 365.25 = 2191.5$),
we also obtain the total signal rate.
To be conservative we computed
the daily rates using the position of Earth on January $1$,
even though the DM flux is expected to be highest in June and \emph{lowest}
in December or January (see e.g.~\cite{Freese:2012xd}).

The dominant background in CYGNUS originates from radioactivity of the
detector and surrounding materials, with a rate estimated to be of
order $10^4$ electron events/keV/year and approximately flat
with energy~\cite{Vahsen:2020pzb}. The energy resolution is of order
$10\%$ at few keV and is anticipated to be 
\begin{equation}
	\frac{\sigma_E}{E} \simeq 10\%\sqrt{\frac{5.9\,{\rm keV}_{ee}}{E}}\,,
\end{equation}
following~\cite{Vahsen:2020pzb} (see also the recent review
of directional DM detection~\cite{Vahsen:2021gnb}).\footnote{For simplicity
  we ignore readout noise, which might lead to further broadening of the
  peak but is not expected to significantly change our conclusions;
  see~\cite{Vahsen:2020pzb} for discussion.}
A significant rate of single photons originating from the decay of
$\chi_2$ might be interpreted as an excess in electron-like events,
and therefore one might expect sensitivity to this signal to require
the total signal rate exceed the expected background rate. 
However, the identifiable pattern of sidereal-daily modulation
should allow for significantly-improved discrimination from background,
as described below.
 
Following our previous work~\cite{Eby:2019mgs}, we perform a simplified
modulation analysis by dividing a sidereal day into two equal-size bins,
labeled ``signal on'' and ``signal off'', corresponding to the half-day
with highest and lowest rates, respectively. 
The reach of CYGNUS can then be estimated by requiring that the 
rate during ``signal-on'', $\Gamma_{\rm on}$, does not exceed the expected
statistical fluctuation of events during ``signal-off'', whose rate is
$\Gamma_{\rm off}$; for a confidence interval of $n\sigma$ we require 
\begin{equation}
 \frac{\Gamma_{\rm signal}}{\Gamma_{\rm off}} \equiv 
 		\frac{\Gamma_{\rm on}-\Gamma_{\rm off}}{\Gamma_{\rm off}} \lesssim \frac{n\times 1.64}{\sqrt{N_{\rm off}}}\,
\end{equation}
where $N_{\rm off} \equiv \Gamma_{\rm off} t_{\rm off}$ and
$t_{\rm off}=(N_{\rm days}/2)$~days is the total time during ``signal-off''.  Note that $\Gamma_{\rm off}$ contains a contribution from both the DM signal, during the ``signal-off'' period, and any additional intrinsic background in the experiment.
The resulting $1\sigma$ sensitivity estimation is given in
Figure~\ref{fig:sensitivity} by the dark-blue region bounded by a
solid blue line. Note that a dedicated search using a more
robust modulation analysis (taking into account the predicted peak shape,
for example) may significantly improve sensitivity relative to
our estimates. 
 
Note that the CYGNUS collaboration have proposed several ways to reduce
their intrinsic backgrounds even further, with the eventual target of
a zero-background experiment~\cite{Vahsen:2020pzb}.
We estimate the resulting sensitivity in this case using a
total integration time of $t_{\rm int}=1$~year; the result is given by
the light-blue shaded regions in Figure~\ref{fig:sensitivity}.
We observe that after only $1$ year of zero-background CYGNUS running,
one would already obtain substantial improvement in sensitivity relative
to either $6$ years of running with expected backgrounds, or existing
direct detection experiments (see Section~\ref{ssec:LZ}). This emphasizes
the importance of further background rejection techniques
in the development of CYGNUS-like experiments.

In Figure~\ref{fig:sensitivity}, we observe that the sensitivity diminishes
at both small and large $g_M$. At small $g_M$, the total signal rate simply
drops and the modulating signal becomes more difficult to observe.
On the other hand, at large $g_M$, the sensitivity is cut abruptly
once the decay length of the excited state becomes shorter than (roughly)
the typical overburden of rock that the DM sees along its path to
the detector. Gran Sasso is located at latitude $43^{\circ}$N and $1.4$~km
depth, and its typical overburden is of order $\ell_{\rm ob}\sim 2$~km,
whereas SUPL is located at $37^{\circ}$S at a depth of $1$~km which
implies a larger average overburden of $\ell_{\rm ob}\sim 10$~km;
see~\cite{Eby:2019mgs} for details about the dependence of overburden
on latitude. Therefore we expect the modulation to be significantly
diminished once the most energetic DM particles have decay lengths
$\ell_{\chi_2}$ shorter than $\ell_{\rm ob}$; we illustrate this using the
gray dashed contours, which match roughly the shape of the blue contours
as expected. 

Note that, as explained above, the sensitivity is also reduced
at small $\delta$ once the decay length becomes comparable to the size
of the Earth; solving Eq.~\eqref{eq:decaylength} for
$\ell_{\chi_2}=R_E$ gives
$g_M\sim10^{-3}(250\,{\rm keV}/\delta)^{3/2}\sqrt{v/450\,{\rm km/sec}}$ for
$m_\chi={\rm TeV}$. However, due to several mitigating factors below
$\delta\sim100\,{\rm keV}$, including the increased range of
kinematically-allowed velocities and possible upscattering targets,
the actual reduction in sensitivity in Figure~\ref{fig:sensitivity}
occurs at $\delta\lesssim30-50\,{\rm keV}$ rather than
$\mathcal{O}(100)$s~keV\@.

Finally, in Figure~\ref{fig:sensitivity} we also observe a distinct difference in the sensitivity contours between the two locations we considered, Gran Sasso and SUPL\@.  
There is a decrease in sensitivity for SUPL when compared to Gran Sasso at large $\delta$, most pronounced for the zero-background sensitivity contour (light blue).  This is due the shape of the modulation signal in SUPL, which is broader than at Gran Sasso; see Figure~\ref{fig:modulationplot}.  This leads to sizable signal rate during the ``signal-off'' time in our simplified on-off analysis.
This decrease is despite the relatively large total signal rate as illustrated by the red contours, and motivates an optimized modulation analysis.
In addition, below the light gray region (discussed in the next section), there is a significant increased sensitivity to modulation at SUPL compared with Gran Sasso, broadly for $\delta \lsim 100$~keV and $g_M \lsim 0.01$.  
Computing the characteristic decay length of the excited state in this region,  we find from Eq.~(\ref{eq:decaylength})
that $\ell_{\chi_2} \gtrsim R_E$
(see also Figure~\ref{fig:decaylength}).  Since SUPL's location in the southern hemisphere implies the dark matter wind is coming from directions that are below the horizon for most of the day, there is a significant increase in the upscatter rate arising from the much larger density of iron in the core of the Earth.  Once the $(\delta,g_M)$ values give shorter decay lengths, the excited state upscattering in the core does not typically live long enough to reach the location of SUPL, causing a return to sensitivities that more closely resemble those at Gran Sasso, which is not able to benefit from the iron core upscatter enhancement in the rates.

\subsection{Constraints from direct detection} \label{ssec:LZ}

Direct detection experiments searching for the nuclear recoil signal arising from the inelastic scattering of
$\chi$ off nuclei are also highly sensitive; we focus here on
LZ~\cite{LZ:2022lsv}, which has the strongest limit to date on the
parameter range considered in this work. 

The rate for the nuclear recoil signal can be obtained by using Eq.~(\ref{eq:dRdER}) and integrating over the various isotopes of xenon with the kinematically-allowed range of velocities permitted by an inelastic collision \cite{Bramante:2016rdh}. The cross section itself, $d\sigma/dE_R$, is proportional to
Eq.~\eqref{eq:dsigmadcostheta}, and for a given xenon isotope is calculable using the product of response functions in
Eq.~\eqref{eq:MNRsqu}. We have used the results
of~\cite{Anand:2013yka}, that provided the nuclear structure
functions for all relevant isotopes of xenon, with mass numbers
$A=128,129,130,131,132,134,$ and $136$. Using the natural abundances
of each isotope~\cite{nuclearwallet} as a proxy for xenon in the detector,
we sum the contributions of each to the rate in Eq.~\eqref{eq:dRdER},
while fixing the total LZ fiducial target mass to
$5.5\,{\rm t}$~\cite{LZ:2022lsv}. The total rate $R$ is then computed
as the integral of $dR/dE_R$ over $E_R$ in the sensitive range of
the detector; for LZ, the detection efficiency is $\gtrsim 50\%$ for
$5\,{\rm keV}\lesssim E_R \lesssim 55\,{\rm keV}$, which we take
as the relevant range for our estimation.\footnote{For simplicity
  we assume $100\%$ efficiency for
  $5\,{\rm keV}\lesssim E_R \lesssim 55\,{\rm keV}$ and $0\%$ elsewhere.}
Finally, we estimate the expected number of events from MIDM in the detector
during the LZ runtime of $t_{\rm int}=60$ live days~\cite{LZ:2022lsv}
as $N_{\rm sig}\simeq R\cdot t_{\rm int}$. 
If $N_{\rm sig}$ exceeds the
statistical fluctuation in the number of observed events in this
energy range, the corresponding parameters are therefore constrained.

The most recent dark matter search results obtained by the LZ experiment yielded $\sim 11$ events
in the nuclear recoil band~\cite{LZ:2022lsv}.  This number corresponds
to the expected background under the assumption that the magnetic dipole transition inelastic nuclear scattering, 
$\chi + \Ncal \rightarrow \chi_2 + \Ncal$, yields an $S1$ and $S2$ response consistent with the LZ experiment's calibrated nuclear recoil band.
Using a $90\%$ confidence interval
around the observed $11$ events gives $N_{\rm sig} \lesssim 16.6$ events;
the resulting constraint in the $g_M-\delta$ plane is shown by the
light-gray shaded region in Figure~\ref{fig:sensitivity}. We observe that,
even in this case, the sensitivity of CYGNUS using current background
estimations is still competitive with direct-detection constraints.
Significant gains are possible for a southern-hemisphere detector
(right panel) and/or if the background estimates improve in the future. 

It is interesting to consider that the LZ nuclear recoil band was
established through neutron scattering off xenon.
The magnetic dipole transition off nuclei proceeds, however, through photon exchange whose cross section
is dominated by the $\mathcal{O}_5$ nuclear response for the recoil energies of relevance to the bound.  This interaction is significantly enhanced  
at low $q^2$, and is quite unlike the contact interaction, $\mathcal{O}_1$, that neutron scattering captures in the calibration of LZ\@.  The central value and 
spread of the nuclear recoil band has not been calibrated to a 
dark matter candidate that may be affected by 
the presence of the electrons.  This suggests that it is
at least possible the $S1$ and $S2$ response obtained from
LZ's neutron calibration may not accurately represent the the
nuclear recoil band for MIDM scattering.
The most conservative interpretation of the LZ data for placing a
limit on MIDM scattering would be to ignore the nuclear recoil band,
and use \emph{all} observed background events to set a limit;
in the recent run, this total was $333$ events 
(see Table 1 of~\cite{LZ:2022lsv}). 
The nuclear recoil calibration-independent bound can be set by
using a $90\%$ confidence interval around $333$ events
(i.e.\ $356$ events, assuming Gaussian distribution)
as shown by the dark gray shaded region in Figure~\ref{fig:sensitivity}.
This is definitively the upper bound for any possible scattering
off xenon that leaves an electronic signal in the detector.  Depending on the details of the $S1$ and $S2$ response
to the specific nuclear responses involved for MIDM scattering,
the true limit will lie somewhere between the dark and light shaded
gray regions.  We suspect that, given the typical momentum exchange  
even at the lowest recoil energies is 
$|q| \simeq \sqrt{2 m_{\rm Xe} E_R} \simeq 30$~MeV  
 (for $E_R \simeq 3$~keV), which is large compared to the inner electron binding energies, 
 the $S1$ and $S2$ response is likely closer
to the neutron scattering band.  However, we believe it is important
to not absolutely dismiss a potential difference, and so we show both the
dark gray and light gray regions
in Figure~\ref{fig:sensitivity} to remind the reader of this subtlety.

\section{Discussion}

We have demonstrated that magnetic inelastic dark matter can be
well-tested by a large underground gaseous detector using
sidereal-daily modulation as the operative method to find
the excited state decay to a photon line signal
and distinguish it from background.
There is clearly an advantage if the detector
is located in the southern hemisphere, e.g., at SUPL, since this
gives the largest rates and the maximum potential for discovery
as we show in Figure~\ref{fig:sensitivity}.
While all elements in the Earth contribute the total rate, we find the
specific elements $^{27}{\rm Al}$ and $^{56}{\rm Fe}$ dominate the
contribution to the upscatter process for the inelasticities
that can be probed: ${\rm few~keV} \, \lesssim \delta \lesssim 150$~keV\@.
Smaller inelasticities $\delta \lesssim {\rm few}$~keV suffer
an increasingly larger fraction of excited states decaying
on distances much larger than the Earth, and thus suppressing
the total rate.  Larger inelasticities $\delta \gtrsim 150$~keV
suffer from a much smaller fraction of elements in the Earth
that permit upscattering, as well as excited states decaying
on distances that can be
small compared with the detector overburden ($\sim 1$~km),
diminishing the modulation of the signal.
The ``sweet spot'' for the photon line
signal that arises from magnetic inelastic dark matter fortuitously
occurs in an energy regime that large gaseous detectors are capable
of detecting.

That said, the existing xenon experimental
results already probe a substantial portion of the parameter space.
Exactly how much parameter space depends in part on whether the
$S1$ and $S2$ response that the xenon experiments have calibrated
using neutron scattering overlaps well with the nuclear recoil
response arising from the magnetic inelastic transition.
The reason for a possible difference has to
do with the dominance of the $\mathcal{O}_5$ nuclear response that is
strongly enhanced at low $q^2$ due to photon exchange,
which is quite unlike the calibration performed with
neutron scattering that is dominated by very short range
(pion exchange and contact) interactions. 
Given that the typical momentum exchange is larger
than $30$~MeV even at the lowest recoil energies that the
experiments measure, we suspect that the xenon detector response
is likely similar to neutron scattering, but we emphasize that this is
an important issue to revisit.

The large gaseous detector that provided our probe of the photon signal
for magnetic inelastic dark matter, CYGNUS, remains at the proposal stage.
Xenon experiments, on the other hand, are well underway probing ever-smaller 
cross sections that include magnetic inelastic dark matter
through upscattering off nuclei.  Our view of the interplay among
these experiments is that \emph{if} xenon experiments were to find
evidence for a nuclear recoil signal above background, the
entire community would be
immediately interested in understanding the nature of this
dark matter signal and its interactions with the SM,  
as well as the astrophysical properties of the local dark matter
halo, such as the local density and velocity distribution.
Our study demonstrates
that large gaseous detectors are \emph{complementary} in their
ability to probe the specific theory that was the focus of this
paper: magnetic inelastic dark matter.  Given that the inelasticity
could be small, and thus may be difficult to discern in the nuclear
recoil energy spectrum itself, our technique of using upscattering off
the elements in the Earth and subsequent decay to a photon would
provide outstanding information to distinguish among the various
possible dark matter theory explanations.

While our main results in Figure~\ref{fig:sensitivity} used the dark matter mass
$m_\chi = 1$~TeV, here let us comment on what happens
as one deviates from this value.  In the case where $m_{\pm} \sim m_\chi$, we can lower $m_\chi$ to a few hundred GeV before running into LHC bounds (and 
at $m_{\pm} = m_\chi = 100$~GeV, the robust LEP II bounds) on charged particles.  In this regime, lowering the dark matter mass will result in the standard increase in the rates from the increase in the dark matter number density holding the local dark matter energy density fixed, i.e., $n_\chi \propto \rho_\chi/m_\chi$.  The second effect is related to the strength of the upscattering and subsequent decay.  These rates, however, depend on only on the magnetic dipole transition coefficient, $\tilde{\mu}_\chi$ (when $m_\chi \gg m_{\Ncal}$, so that the reduced mass that enters the upscatter cross section is well-approximated by just the nucleus mass),
that has the mass dependence $\tilde{\mu}_\chi \propto 1/m_\chi$.  Combining these effects gives the approximate scaling of the rates, $\sigma_\chi n_\chi v \propto g_M^2/m_\chi^3$.  One can therefore read off the bounds and sensitivities to $g_M$ for other dark matter masses $m_\chi \gtrsim \mathcal{O}(100s)$~GeV by reinterpreting the $y$-axis to be 
\begin{eqnarray}
g_M(m_\chi) &=& g_M(1 \, {\rm TeV}) \left( \frac{m_\chi}{1 \; {\rm TeV}} \right)^{3/2} \, ,
\end{eqnarray}
i.e., probing smaller (larger) values of $g_M$ for smaller (larger) masses $m_\chi$.  If instead we imagine $m_{\pm} \gg m_\chi$, then the $y$-axis becomes
\begin{eqnarray}
g_M(m_\pm,m_\chi) &=& g_M(1 \, {\rm TeV}, \, 1 \, {\rm TeV}) \left( \frac{m_\chi}{1 \; {\rm TeV}} \right)^{1/2} \left( \frac{m_\pm}{1 \; {\rm TeV}} \right) \, , 
\end{eqnarray}
where this expression continues to assume $m_\chi \gg m_{\Ncal}$.
If instead we consider changing $m_\chi$ to be near or below $m_{\Ncal}$, there are several nonlinearities that prevent a simple scaling:  
there is a reduction in the rate due to the dependence of the reduced mass on $m_\chi$, and the sidereal daily modulation is 
diminished significantly since the scattering is no longer dominantly in the forward direction.

One of the auxiliary results
from our analysis is that we also bound the magnetic dipole moment
of dark matter (in the limit $\delta \ra 0$); the limit can be readily obtained from Figure~\ref{fig:sensitivity}.
For the purposes of this paragraph,
let's write the magnetic dipole moment as in Eq.~\eqref{eq:dipolemoment}, i.e.
$\tilde{\mu}_\Psi = g_M e/(4 m_\chi)$,
using the same dimensionless coefficient $g_M$.
Then, using the bound we obtained from 
the LZ nuclear recoil band (i.e., assuming this applies
for a magnetic dipole moment interaction), we find
\begin{eqnarray}
g_M & \lesssim & 0.0026 \qquad [m_\chi \, = \, 1 \, {\rm TeV}] \, .
\end{eqnarray}
The direct detection bound scales proportional to
$\sigma_\chi n_\chi \propto g_M^2/m_\chi^3$ 
when $m_\chi \gg m_\Ncal$.  
For an electroweak dark matter candidate
that acquires a magnetic dipole moment through a loop of charged particles
(e.g., a one-loop diagram with $W$ and a set of electrically
charged excited states in the dark sector), we estimate
$g_M \sim \frac{g^2 m_\chi}{16 \pi^2 m_{\pm}}$, where
$g \simeq 0.65$ is the electroweak coupling.  In the case where
the charged states in the dark sector are only slightly heavier
than the dark matter itself, $m_{\pm} \simeq m_\chi$, the
LZ bound corresponds to 
\begin{eqnarray}
m_\chi \; >& 1.0 \; {\rm TeV} &\qquad [g_M = g^2/(16 \pi^2)] \, .
\end{eqnarray}
In the case of strongly-coupled dark matter,
for two cases of $g_M$ we obtain 
\begin{eqnarray}
  m_\chi \; >&  52 \; {\rm TeV} &\qquad [g_M = 1] \, , \\
  m_\chi \; >& 126 \; {\rm TeV} &\qquad [g_M = g_{\rm neutron} = -3.83] \, .
\end{eqnarray}
It is striking that the xenon experiments are now probing
dark matter masses over $100$~TeV when dark matter has a
magnetic dipole moment comparable to the neutron!

We also have calculated the nuclear responses for a much larger
set of elements than originally presented in \cite{Fitzpatrick:2012ix}
by using the {\bigstick} nuclear physics code.  
Some of our results are new, while others we were able to
cross-check with Refs.~\cite{Fitzpatrick:2012ix,Catena:2015uha};
we find agreement in all of the overlapping results (see Figure~\ref{fig:Comparison}).
We have provided all of our numerical results for the
response functions as auxiliary files in~\cite{Eby_IsotopeResponses}.

Finally, we remark that there is an interesting parameter
space that may provide an opportunity for xenon experiments
themselves to be sensitive to the photon line signal.
Once the dark matter mass is small, $m_\chi \lesssim 10$~GeV,
the nuclear recoil constraints become rapidly weaker because
nuclear scattering leaves an energy deposition in the detector
that is nearly always below the energy threshold for detection.
However, for $1\,{\rm GeV} \lesssim m_\chi \lesssim 10$~GeV, it is possible
for magnetic inelastic dark matter to still have a substantial rate
to upscatter in the Earth, with the excited state decaying
into a photon line signal.  So long as
${\rm few~keV} \lesssim \delta \lesssim 10$~keV,
the photon line would show up as part of the \emph{electron
  background}.  While there is unfortunately only a very small
modulating event fraction, the total event rate is,
nevertheless, potentially large enough that this could be
seen above backgrounds.
Hence, our signal could provide a 
unique opportunity to access magnetic inelastic dark matter 
using existing xenon experiment data, 
that we plan to return to in future work \cite{Eby:2024toappear}.

\section*{Acknowledgments}

We thank W.~Haxton for help with {\bigstick},
D.~Snowden-Ifft for discussions of DRIFT and S.~Vahsen
for discussions of CYGNUS\@.
The work of JE was supported by the
World Premier International Research Center Initiative (WPI), MEXT, Japan and
by the JSPS KAKENHI Grant Numbers 21H05451 and 21K20366, as well as by the
Swedish Research Council (VR) under grants 2018-03641 and 2019-02337.  
The work of PJF was supported by the DoE under contract number DE-SC0007859
and Fermilab, operated by Fermi Research Alliance, LLC under contract
number DE-AC02-07CH11359 with the United States Department of Energy.
The work of GDK was supported in part by the
U.S. Department of Energy under Grant Number DE-SC0011640.

\appendix

\section{Electromagnetic Moments and Transitions}
\label{app:emoperators}
\numberwithin{equation}{section}
\setcounter{equation}{0}

In this appendix we clarify the allowed operators and interactions
in the presence of one or more massive fermions,
following the notation of \cite{Dreiner:2008tw}. 
As is well-known, a single Weyl fermion $\xi$
transforms under a continuous $U(1)$ flavor symmetry,
$\xi \ra e^{i \theta} \xi$, that is explicitly broken
by a (Majorana) mass term $m_\xi \xi \xi + h.c.$
Moreover, a single Weyl fermion cannot have a magnetic dipole moment
since $\xi \sigma^{\mu\nu} \xi = -\xi \sigma^{\mu\nu} \xi$
has only the trivial solution $\xi = 0$.  
This follows since two anti-communting fermions $\xi$, $\chi$ satisfy
\begin{eqnarray}
\xi \sigma^{\mu\nu} \chi
  &=& -\chi \sigma^{\mu\nu} \xi 
\label{eq:LHsigmamunuidentity}
\end{eqnarray}
where
\begin{eqnarray}
\sigma^{\mu\nu}
  &\equiv& \frac{i}{4}
      \left( \sigma^\mu \bar{\sigma}^\nu - \sigma^\nu \bar{\sigma}^\mu \right)
\end{eqnarray}
and $\sigma^\mu = ( 1, \sigma_i)$, $\bar{\sigma}^\mu = ( 1, -\sigma_i)$, 
with $\sigma_i$ the usual Pauli sigma matrices.  
The Hermitian conjugate of Eq.~(\ref{eq:LHsigmamunuidentity}) is
\begin{eqnarray}
\chi^\dagger \bar{\sigma}^{\mu\nu} \xi^\dagger
  &=& -\xi^\dagger \bar{\sigma}^{\mu\nu} \chi^\dagger
\end{eqnarray}
with
\begin{eqnarray}
\bar{\sigma}^{\mu\nu}
  &\equiv& \frac{i}{4}
      \left( \bar{\sigma}^\mu \sigma^\nu - \bar{\sigma}^\nu \sigma^\mu \right)
      \, .
\end{eqnarray}

Given two Weyl fermions $\xi$ and $\chi$, we can write the
dimension-5 interaction, with coefficient $\tilde{\mu} + i \tilde{d}$,
where $\tilde{\mu}$ and $\tilde{d}$ are real, as
\begin{eqnarray}
  \left( \tilde{\mu} + i \tilde{d} \right) 
  \chi \sigma^{\mu\nu} \xi F_{\mu\nu} + h.c.
  &=& \left[ \left( \tilde{\mu} + i \tilde{d} \right)
      \chi \sigma^{\mu\nu} \xi + h.c. \right] F_{\mu\nu}
      \nonumber\\
  &=& \tilde{\mu} \left( \chi \sigma^{\mu\nu} \xi
                  + \xi^\dagger \bar{\sigma}^{\mu\nu} \chi^\dagger \right) 
                  F_{\mu\nu}
  + i \tilde{d} \left( \chi \sigma^{\mu\nu} \xi
                  - \xi^\dagger \bar{\sigma}^{\mu\nu} \chi^\dagger \right) 
                  F_{\mu\nu}
                  \, .
\label{eq:twocomponentmagneticmoment}
\end{eqnarray}
If $\xi$ and $\chi$ have a Dirac mass
\begin{eqnarray}
m_D \, \xi \chi + h.c.
\end{eqnarray}
we can group the two Weyl spinors into a single 4-component Dirac spinor
\begin{eqnarray}
\psi_D &\equiv& { \xi \choose \chi^\dagger } \, ,
\end{eqnarray} 
and then Eq.~(\ref{eq:twocomponentmagneticmoment}) becomes
\begin{eqnarray}
\left( \tilde{\mu} + i \tilde{d} \right) 
  \chi \sigma^{\mu\nu} \xi F_{\mu\nu} + h.c.
  &=& \frac{\tilde{\mu}}{2} \bar{\psi}_D \Sigma^{\mu\nu} \psi_D F_{\mu\nu}
      -i \frac{\tilde{d}}{2} \bar{\psi}_D \Sigma^{\mu\nu} \gamma_5 \psi_D
         F_{\mu\nu}
\label{eq:magneticmoment}
\end{eqnarray}
where
\begin{eqnarray}
\Sigma^{\mu\nu} &\equiv& \frac{i}{2} \left[ \gamma^\mu, \gamma^\nu \right] 
  \; = \; 2 \left( \begin{array}{cc} \sigma^{\mu\nu} & 0 \\
                                     0 & \bar{\sigma}^{\mu\nu} 
                   \end{array} \right) \, .
\label{eq:sigmamunu}
\end{eqnarray}
Eq.~(\ref{eq:magneticmoment}) represents, of course,
simply the magnetic and electric dipole moments
of a single Dirac fermion.

\subsection{Electromagnetic Transition Operators with two Weyl Fermions}

A simple construction of a magnetic dipole transition operator can
be obtained by starting with two Weyl fermions that have
a Dirac mass and the magnetic dipole moment
interaction Eq.~(\ref{eq:magneticmoment}) \cite{Kopp:2009qt,Chang:2010en}.
Next, introduce Majorana masses for the Weyl fermions
(and for simplicity of the presentation below, we set these equal)
\begin{eqnarray}
\frac{m_M}{2} \, ( \xi \xi + h.c. ) + \frac{m_M}{2} \, ( \chi \chi + h.c.) \, ,
\end{eqnarray}
and the mass matrix for the Weyl fermions becomes
\begin{eqnarray}
\frac{1}{2} \left( \xi \;\; \chi \right)
  \left( \begin{array}{cc} m_M & m_D \\
                           m_D & m_M \end{array} \right)
   { \xi \choose \chi } + h.c. \, .
\end{eqnarray}
The Weyl fermions can be diagonalized into mass eigenstates $\psi_1, \psi_2$
using 
\begin{eqnarray}
  \xi  &=& \frac{1}{\sqrt{2}} \left( \psi_2 + i \psi_1\right) \,, \\
  \chi &=& \frac{1}{\sqrt{2}} \left( \psi_2 - i \psi_1 \right) \,,
\label{eq:fermiondiagonalization}
\end{eqnarray}
resulting in
\begin{eqnarray}
  \frac{1}{2}
  (m_D - m_M) (\psi_1 \psi_1 + h.c.)
  + \frac{1}{2} (m_D + m_M) (\psi_2 \psi_2 + h.c.) \, .
\end{eqnarray}
The would-be magnetic dipole moment interaction becomes a magnetic
dipole transition interaction
\begin{equation}
  \tilde{\mu} (\chi \sigma^{\mu\nu} \xi + h.c.) F_{\mu\nu} \; \longrightarrow \;
  \tilde{\mu} (-i \psi_1 \sigma^{\mu\nu} \psi_2 + h.c.) F_{\mu\nu}
  = -i \tilde{\mu} \left(\psi_1 \sigma^{\mu\nu} \psi_2
              + \psi_1^\dagger \bar{\sigma}^{\mu\nu} \psi_2^\dagger \right)
              F_{\mu\nu} 
              \, ,
\end{equation}
where the second term
has a relative minus sign, i.e., it is equal to 
$- \psi_2^\dagger \bar{\sigma}^{\mu\nu} \psi_1^\dagger$, 
because the Weyl fermion mass eigenstate diagonalization,
Eq.~(\ref{eq:fermiondiagonalization}),
involves $\pm i \psi_1$ and not $\pm \psi_1$.

Defining the Majorana four-component fermions with
\begin{eqnarray}
\Psi_{M1} \; \equiv \; { \psi_1 \choose \psi_1^\dagger }
  &\quad&
\Psi_{M2} \; \equiv \; { \psi_2 \choose \psi_2^\dagger }
\end{eqnarray}
allows the rewriting of magnetic dipole transition operator into
\begin{eqnarray}
\tilde{\mu} (-i \psi_1 \sigma^{\mu\nu} \psi_2 + h.c.) F_{\mu\nu}
  &=& -i \frac{\tilde{\mu}}{2} \bar{\Psi}_{M1} \Sigma^{\mu\nu} \Psi_{M2}
      F_{\mu\nu} \, ,
\label{eq:majoranamagnetictransition}
\end{eqnarray}
where we note that
$\bar{\Psi}_{M1} \Sigma^{\mu\nu} \Psi_{M2} =
-\bar{\Psi}_{M2} \Sigma^{\mu\nu} \Psi_{M1}$ 
and hence Eq.~(\ref{eq:majoranamagnetictransition})
is manifestly Hermitian, in one-to-one correspondence with the
manifestly-Hermitian magnetic dipole moment operator
from which it was derived. 
Similarly, the electric dipole transition operator is
\begin{eqnarray}
  (i \tilde{d} \, \chi \sigma^{\mu\nu} \xi + h.c.) F_{\mu\nu}
  \; \longrightarrow \;
  \tilde{d} (-\psi_1 \sigma^{\mu\nu} \psi_2 + h.c.) F_{\mu\nu}
  &=& - \tilde{d} \left(\psi_1 \sigma^{\mu\nu} \psi_2
              + \psi_2^\dagger \bar{\sigma}^{\mu\nu} \psi_1^\dagger \right)
              F_{\mu\nu} \nonumber \\
  &=& \frac{\tilde{d}}{2} \, \bar{\Psi}_{M1} \Sigma^{\mu\nu} \gamma_5 \Psi_{M2}
      F_{\mu\nu} \, ,
\end{eqnarray}
and we note 
$\bar{\Psi}_{M1} \Sigma^{\mu\nu} \gamma_5 \Psi_{M2} =
\bar{\Psi}_{M2} \Sigma^{\mu\nu} \gamma_5 \Psi_{M1}$.

\subsection{Electromagnetic Transition Operators with two Dirac Fermions}

Another possibility for electromagnetic dipole transitions is
between two Dirac fermions 
\begin{eqnarray}
  \Psi_1 \; \equiv \; { \xi_1 \choose \chi_1^\dagger } &\quad&
  \Psi_2 \; \equiv \; { \xi_2 \choose \chi_2^\dagger } \, ,
\end{eqnarray}
with different masses $m_1 < m_2$
\begin{eqnarray}
  m_1 (\xi_1 \chi_1 + h.c.) + m_2 (\xi_2 \chi_2 + h.c.)
  &=& m_1 \bar{\Psi}_1 \Psi_1 + m_2 \bar{\Psi}_2 \Psi_2 
  \, , 
\end{eqnarray}
that explicitly break the flavor $U(2)$ symmetry down to
$U(1)_1 \times U(1)_2$.  
A general set of electromagnetic dipole moments and dipole transition
operators are
\begin{eqnarray}
\left( \chi_1 \;\; \chi_2 \right)
  \left( \begin{array}{cc}
   \tilde{\mu}_{11} + i \tilde{d}_{11} & \tilde{\mu}_{12} + i \tilde{d}_{12} \\ 
   \tilde{\mu}_{21} + i \tilde{d}_{21} & \tilde{\mu}_{22} + i \tilde{d}_{22}
  \end{array} \right)
\sigma^{\mu\nu} { \xi_1 \choose \xi_2 } F_{\mu\nu} + h.c. \, ,
\end{eqnarray}
and in the Dirac basis, 
\begin{eqnarray}
\frac{1}{2}
\left( \bar{\Psi}_1 \;\; \bar{\Psi}_2 \right)
  \left( \begin{array}{cc}
   \tilde{\mu}_{11} & \tilde{\mu}_{12} \\ 
   \tilde{\mu}_{21} & \tilde{\mu}_{22} 
  \end{array} \right)
\Sigma^{\mu\nu} { \Psi_1 \choose \Psi_2 } F_{\mu\nu} + h.c. \\
-\frac{i}{2}
\left( \bar{\Psi}_1 \;\; \bar{\Psi}_2 \right)
  \left( \begin{array}{cc}
   \tilde{d}_{11} & \tilde{d}_{12} \\ 
   \tilde{d}_{21} & \tilde{d}_{22}
  \end{array} \right)
\Sigma^{\mu\nu} \gamma_5 { \Psi_1 \choose \Psi_2 } F_{\mu\nu} + h.c. \, .
\end{eqnarray}  
Here we see both magnetic and electric dipole moments
(proportional to $\tilde{\mu}_{ii}$ and $\tilde{d}_{ii}$)  
as well as magnetic and electric dipole transitions 
(proportional to $\tilde{\mu}_{12}$,$\tilde{\mu}_{21}$ and
$\tilde{d}_{12}$,$\tilde{d}_{21}$).
The dipole moments obviously preserve $U(1)_1$ and $U(1)_2$ separately,
while the dipole transition operators explicitly violate the
combination $U(1)_1 + U(1)_2$ while leaving $U(1)_1 - U(1)_2$ intact.

Focusing on the magnetic dipole moments and transitions,
we emphasize that the coefficients $\tilde{\mu}_{ij}$ 
are independent.  Transitions between $\Psi_1$ to $\bar{\Psi}_2$
occur independently of $\bar{\Psi}_1$ to $\Psi_2$.
For example, in a universe in which dark matter is $\Psi_1$,
and there was an asymmetric abundance mechanism that
populated only $\Psi_1$-type particles (and not the
$\bar{\Psi}_1$-type anti-particles),
the allowed transitions are between $\Psi_1$
to $\bar{\Psi}_2$ and then the decay $\bar{\Psi}_2$
to $\Psi_1$.  In theories that preserve $\mathrm{CP}$, 
the coefficients $\tilde{\mu}_{12} = \tilde{\mu}_{21}$,
but this does not affect the independent nature of these
transitions.

\section{Nuclear Response and the One-Body Density Matrix}
\label{app:OBME}

The determination of the nuclear response functions $W_k$ requires
knowing the nuclear wave functions since they are expectation values
of operators in nuclear states,
\begin{eqnarray}
W_{\Ocal^{A}\Ocal^{B}}^{\t\t'}
  &=& \sum_{J=0}^\infty \bra{j_\nucleus} | \Ocal^{A}_{J;\t} | \ket{j_\nucleus}
      \bra{j_\nucleus} | \Ocal^{B}_{J;\t'} | \ket{j_\nucleus} \, .
\end{eqnarray}
The many-body matrix element of a one-body operator (as above)
can be written as a product of the one-body density matrix times
the matrix elements of the one-body operator:
\begin{equation}
\bra{j_\nucleus; T\,M_\nucleus} | \sum_{i=1}^A\Ocal_{J;\t}(q\,\vec{x}_i) |
  \ket{j_\nucleus; T\,M_\nucleus}
  = (-1)^{T-M_T} \left( \begin{matrix}
					 T & \t & T \\
				  -M_T & 0  & M_T
                \end{matrix} \right)
              \bra{j_\nucleus;T} \vdots\vdots
              \sum_{i=1}^A\Ocal_{J;\t}(q\vec{x}) \vdots\vdots \ket{j_\nucleus;T}
\end{equation}
where we have introduced the Wigner-$3j$ symbols (in brackets), and 
\begin{eqnarray}\label{eq:app-obdmO}
\bra{j_\nucleus;T} \vdots\vdots \sum_{i=1}^A\Ocal_{J;\t}(q\vec{x}) \vdots\vdots \ket{j_\nucleus;T} 
				&=& \sum_{|\a|,|\b|} \Psi_{|\a|,|\b|}^{J;\t} \bra{|\a|} \vdots\vdots \Ocal_{J;\t}(q\vec{x}) \vdots\vdots \ket{|\b|}~,
\end{eqnarray}
see Eq.~(61) of \cite{Anand:2013yka}. In the above expressions,
$T$ ($M_T$) is total ($z$-component) of isospin and $\a,\b$ represents
nonmagnetic quantum numbers. The $\vdots \vdots \Ocal \vdots \vdots$
denotes the doubly-reduced matrix element (in spin and isospin), and $\Psi_{|\a|,|\b|}^{J;\t}$ are the One-Body Density Matrix (OBDM)
elements for the ground-to-ground state transition.  The composite
nuclear states are represented by the collective quantum numbers
$\a$ and $\b$.  We assume the underlying single-particle basis is given by states of the harmonic oscillator, in which case, 
\[\ket{\a} = \ket{n_\a(\ell_\a s_\a = 1/2)j_\a m_{j_\a}; t_\a = 1/2 m_{t_\a}}\,.
\] 
Then
\begin{eqnarray}
 \bra{|\a|} \vdots\vdots \Ocal_{J;\t}(q\vec{x}) \vdots\vdots \ket{|\b|} &= \sqrt{2(2\t + 1)} \bra{n_\a(\ell_\a 1/2)j_\a} \left| O_J(q\vec{x}) \right| \ket{n_\b(\ell_\b 1/2)j_\b}  \\
 				&= \frac{\sqrt{2(2\t + 1)}}{\sqrt{4\pi}}\,y^{(J-K)/2}\,e^{-y}\,p(y)
\end{eqnarray}
where $y = (q\,b/2)^2$.  Here $b$ is the oscillation parameter with
\begin{equation}
b=\sqrt{\frac{41.467}{45\,A^{-1/3} - 25\,A^{-2/3}}} \,\rm{ fm},
\end{equation}
and $A$ the mass number in the target.

Therefore, the problem of calculating the matrix elements $\bra{\a}\Ocal\ket{\b}$ reduces to the determination of a finite set of polynomials $p(y)$. 
The quantum numbers on which $p(y)$ depends are:
\begin{itemize}
 \item $N_\a$: Initial and final principal quantum number, $N_\a = 2(n_\a-1) + \ell_\a = (0,1,2,...)$.
 \item $j_\a$: Single-particle wavefunction angular momentum. This determines $\ell_\a$ by the requirement that $(N-\ell)$ be even, i.e. that $n = (N-\ell)/2 + 1 \in \mathbb{Z}$.
 \item $J$: The rank of the operator in the multipole expansion.
 \item $K$ is determined by the parity of the operator: $K=2$ for normal parity (e.g. $M_J$, $\tilde{\Phi}'_J$, and $\tilde{\Phi}''_J$), and $K=1$ for abnormal parity  operators (e.g. $\Delta$, $\Sigma'$, and $\Sigma''$).
\end{itemize}
In \cite{Donnelly:1979ezn}, it is shown that these polynomials have the form
\begin{eqnarray}
 p(y) &=& \sqrt{J_1}\frac{J_2}{J_3}\left[\frac{I_1}{K_1} + \frac{I_2}{K_2}\,y + ... + \frac{I_{n+1}}{K_{n+1}}\,y^n\right] \nonumber \\
 				&\equiv& \sqrt{J_1} \,c\,\left[\eta_0 + \eta_1\,y + ... + \eta_n\,y^n\right]~,
\end{eqnarray}
where the $J_i, K_i, I_i$ are integers.  We need only determine the coefficients in these polynomials once for each operator we use, as they do not depend on the element except in the selection of quantum numbers.  We do so by using the Mathematica package described in Ref. \cite{Anand:2013yka}.

The One Body Density Matrices, the $\Psi$ in Eq.~(\ref{eq:app-obdmO}), can be determined for each element and each quantum number by using a model for the nuclear physics interactions. To do this  we employ a modern automated code, {\bigstick} \cite{Johnson:2018hrx}, which in turn uses input files from other codes, Oxbash \cite{Oxbash} and NuShell \cite{NuShell}. The calculation of the OBDM elements in {\bigstick} requires two files: a model file (typically with extension .sps) and an interaction file (typically with extension .int). 
The model files represent a model for a nuclear core with filled energy levels, plus valence protons and neutrons. In the model sp.sps, for example, the core is an ${}^{16}$O (8 protons and 8 neutrons), corresponding the the magic number $Z=N=8$ of filled states for both the proton and neutron. This corresponds to filling of the lowest-lying orbitals: $1s_{1/2}$ (2 states), with orbital angular momentum $\ell=0$ and total $j=1/2$; plus the lowest-lying $\ell=1$ orbitals (6 total states) labeled $0p_{1/2}$ ($j_z=1/2$) and $0p_{3/2}$ ($j_z=3/2$). The next available states are $0d_{5/2}$, $0d_{3/2}$, and $1s_{1/2}$ (total of 10+2 = 12 states) for the valence protons and neutrons. Once these are all filled, we hit the next magic number $Z=N=20$, fill a new core (now Ca) and begin again with the next set of orbitals. 

The interaction files quantify the interactions among the valence nucleons in the background of the potential in the filled core (e.g. ${}^{16}$O), and so a given interaction model is only relevant in the context of a particular model of the core. 
We tabulate several model/interaction file combinations which are available for use in {\bigstick}, along with their corresponding valence states and range of applicability in $N$ and $Z$, in Table~\ref{tab:models}.
Note that {\bigstick} does not support different interaction models for protons and neutrons and above $Z=20$ the number of protons and neutrons in stable isotopes become increasingly unequal.  Thus, we expect the errors to grow as we move to heavier nuclei.
We combine the polynomials for the matrix elements and the one body density matrices using the Mathematica package of Ref.~\cite{Anand:2013yka} to determine the final nuclear response functions, $W_k$.  Note that doing so requires a little translation to convert the output of {\bigstick} to the input of the Mathematica package. 

\begin{figure}[t]
\centering
\includegraphics[height=5.5cm]{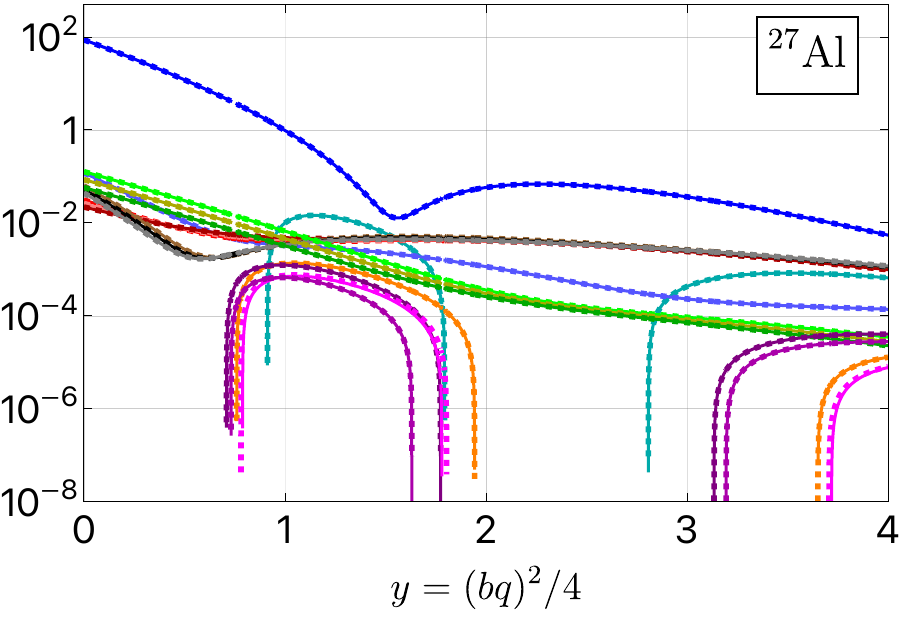}
\,
\includegraphics[height=5.5cm]{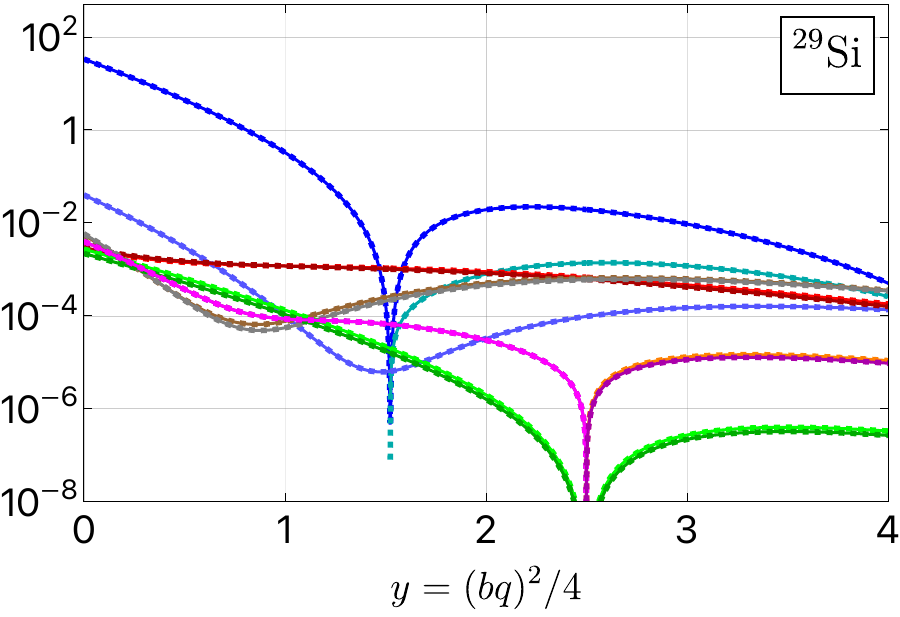}
\\
\includegraphics[height=5.5cm]{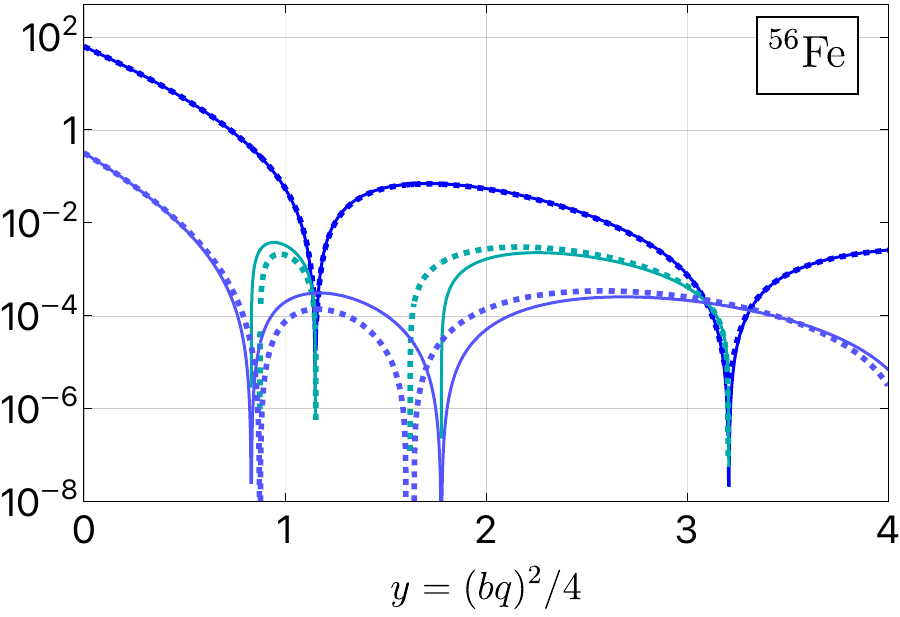}
\, 
\includegraphics[height=5cm]{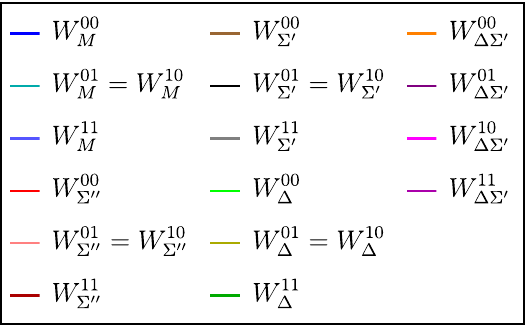}
\caption{Direct comparison of the response functions
  $W_\mathcal{O}^{\tau\tau'}$ from our calculation
  (solid lines) to previous works (dashed lines).
  For $^{27}$Al (upper-left panel), we compare to Ref.~\cite{Catena:2015uha},
  while for $^{29}$Si (upper-right) and $^{56}$Fe (lower-left),
  we compare to Ref.~\cite{Fitzpatrick:2012ix}; line colors correspond
  to the response functions for the five relevant MIDM operators for
  all isospin combinations (see legend).} 
\label{fig:Comparison}
\end{figure}

For some elements and response functions we can compare the output of our approach to results presented in Refs.~\cite{Fitzpatrick:2012ix, Catena:2015uha}; see Figure~\ref{fig:Comparison}.

\begin{table}[t]
\centering
\begin{tabular}{c | c || c | c | c }
 \hline
    model & interaction  & core & Valence States & N , Z range \\
  \hline
  spsdpf  & spsdpfpb & - & 
  \begin{tabular}{c}
  $0s_{1/2}$, $0p_{3/2}$, $0p_{1/2}$, $0d_{5/2}$, $0d_{3/2}$, $1s_{1/2}$, \\
  $0f_{7/2}$, $0f_{5/2}$, $1p_{3/2}$, $1p_{1/2}$
  \end{tabular}
 & $1 \leq N,Z \leq 40$  \\
  p  & pjt  &${}^2$He & $0p_{3/2}$, $0p_{1/2}$ & $2 \leq N,Z \leq 8$ \\
  sdpf & sdpfnow & ${}^{16}$O & 
  \begin{tabular}{c}
    $0d_{5/2}$, $0d_{3/2}$, $1s_{1/2}$, \\
    $0f_{7/2}$, $0f_{5/2}$, $1p_{3/2}$, $1p_{1/2}$ 
  \end{tabular}
  & $8 \leq N,Z \leq 40$ \\
  sd & w & ${}^{16}$O & $0d_{5/2}$, $0d_{3/2}$, $1s_{1/2}$ & $8 \leq N,Z \leq 20$  \\
  pf & gx1 & ${}^{40}$Ca & $0f_{7/2}$, $0f_{5/2}$, $1p_{3/2}$, $1p_{1/2}$ & $20 \leq N,Z \leq 40$ \\
  n50j & n50j & ${}^{56}$Ni & $0f_{5/2}$, $1p_{3/2}$, $1p_{1/2}$, $0g_{9/2}$, $0g_{7/2}$, $1d_{5/2}$ & $28 \leq N,Z \leq 50$ 
  \\
  n82 & n82k & ${}^{100}$Sn & $0g_{9/2}$, $1d_{5/2}$, $2s_{1/2}$, $1d_{3/2}$, $0h_{11/2}$ & $50\leq N,Z \leq 82$ \\
  \hline
\end{tabular}
\caption{A summary of {\bigstick} models/interactions which can be used, along with their range of applicability in $Z$ and $N$. We list all the available options that (a) treat $p$ and $n$ symmetrically, and (b) do not skip over any state (valence states are ordered by energy).}
\label{tab:models}
\end{table}

\subsection{Nuclear Response Functions}
\label{app:Wofp}

Below we present the form of the nuclear response functions for the four operators relevant for MIDM, for each of the elements studied in this paper.  Note that higher-precision coefficients, including all the nuclear response functions ($W_M$, $W_{\Sigma ''}$, $W_{\Sigma '}$, $W_{\Phi''}$, $W_{\tilde{\Phi}'}$, $W_{\Delta}$, $W_{\Phi''M}$, $W_{\Delta\Sigma '}$), for these and additional elements are collected in an ancillary data file on GitHub~\cite{Eby_IsotopeResponses}.  The file contains information for the elements $^{12}$C,$^{14}$N, $^{24}$Mg, $^{25}$Mg, $^{27}$Al, $^{28}$Si, $^{29}$Si, $^{30}$Si, $^{54}$Fe, $^{56}$Fe, $^{57}$Fe, $^{58}$Fe, $^{58}$Ni, $^{60}$Ni, $^{70}$Ge, $^{72}$Ge, $^{73}$Ge, $^{74}$Ge, $^{87}$Sr, and $^{88}$Sr.
We expect the nuclear physics uncertainties to be sub-percent for light elements ($Z\lsim 20$) and grow to $\mathcal{O}(10\%-50\%)$ for elements as heavy as iron, and beyond.

$\mathbf{_{12}^{24}Mg}$
\begin{eqnarray*}
W_M^{00}&=& e^{-2 y} \left(0.123 \left(9.63-7.49 y+ y^2\right)^2\right)\nn 
\end{eqnarray*}

$\mathbf{_{12}^{25}Mg}$
\begin{eqnarray*}
W_M^{00}&=& e^{-2 y} \left(74.6-119. y+65.2 y^2-13.9 y^3+1.07 y^4\right)\nn \\ 
W_M^{10}=W_M^{01}&=& e^{-2 y} \left(-2.98+6.37 y-4.21 y^2+1.01 y^3-0.0761 y^4\right)\nn \\ 
W_M^{11}&=& e^{-2 y} \left(0.119-0.318 y+0.293 y^2-0.0971 y^3+0.0106 y^4\right)\nn \\ 
W_{\Sigma ''}^{00}&=& e^{-2 y} \left(0.0355-0.0509 y+0.0342 y^2-0.0069 y^3+0.00984 y^4\right)\nn \\ 
W_{\Sigma ''}^{10}=W_{\Sigma ''}^{01}&=& e^{-2 y} \left(-0.0306+0.0407 y-0.0237 y^2+0.00464 y^3-0.00902 y^4\right)\nn \\ 
W_{\Sigma ''}^{11}&=& e^{-2 y} \left(0.0264-0.0324 y+0.0162 y^2-0.00299 y^3+0.00837 y^4\right)\nn \\ 
W_{\Sigma '}^{00}&=& e^{-2 y} \left(0.0711-0.233 y+0.253 y^2-0.0904 y^3+0.0214 y^4\right)\nn \\ 
W_{\Sigma '}^{10}=W_{\Sigma '}^{01}&=& e^{-2 y} \left(-0.0613+0.204 y-0.223 y^2+0.0814 y^3-0.0199 y^4\right)\nn \\ 
W_{\Sigma '}^{11}&=& e^{-2 y} \left(0.0529-0.179 y+0.198 y^2-0.0738 y^3+0.0186 y^4\right)\nn \\ 
W_{\Delta }^{00}&=& e^{-2 y} \left(0.123-0.0983 y+0.0214 y^2\right)\nn \\ 
W_{\Delta }^{10}=W_{\Delta }^{01}&=& e^{-2 y} \left(-0.0807+0.0646 y-0.0136 y^2\right)\nn \\ 
W_{\Delta }^{11}&=& e^{-2 y} \left(0.053-0.0424 y+0.00872 y^2\right)\nn \\ 
W_{\Delta \Sigma '}^{00}&=& e^{-2 y} \left(-0.0935+0.191 y-0.0995 y^2+0.0148 y^3\right)\nn \\ 
W_{\Delta \Sigma '}^{10}&=& e^{-2 y} \left(0.0614-0.125 y+0.0641 y^2-0.00934 y^3\right)\nn \\ 
W_{\Delta \Sigma '}^{01}&=& e^{-2 y} \left(0.0806-0.169 y+0.0894 y^2-0.0136 y^3\right)\nn \\ 
W_{\Delta \Sigma '}^{11}&=& e^{-2 y} \left(-0.0529+0.111 y-0.058 y^2+0.00874 y^3\right)\nn 
\end{eqnarray*}

$\mathbf{_{13}^{27}Al}$
\begin{eqnarray*}
W_M^{00}&=& e^{-2 y} \left(87-146 y+83.5 y^2-18.6 y^3+1.43 y^4\right)\nn \\ 
W_M^{10}=W_M^{01}&=& e^{-2 y} \left(-3.22+7. y-4.93 y^2+1.34 y^3-0.115 y^4\right)\nn \\ 
W_M^{11}&=& e^{-2 y} \left(0.119-0.318 y+0.337 y^2-0.133 y^3+0.0182 y^4\right)\nn \\ 
W_{\Sigma ''}^{00}&=& e^{-2 y} \left(0.0309-0.0367 y+0.0265 y^2-0.00242 y^3+0.011 y^4\right)\nn \\ 
W_{\Sigma ''}^{10}=W_{\Sigma ''}^{01}&=& e^{-2 y} \left(0.026-0.0211 y+0.0159 y^2+0.000606 y^3+0.0106 y^4\right)\nn \\ 
W_{\Sigma ''}^{11}&=& e^{-2 y} \left(0.0219-0.00944 y+0.0115 y^2+0.000954 y^3+0.0105 y^4\right)\nn \\ 
W_{\Sigma '}^{00}&=& e^{-2 y} \left(0.0619-0.211 y+0.244 y^2-0.0943 y^3+0.0244 y^4\right)\nn \\ 
W_{\Sigma '}^{10}=W_{\Sigma '}^{01}&=& e^{-2 y} \left(0.052-0.187 y+0.233 y^2-0.0985 y^3+0.0259 y^4\right)\nn \\ 
W_{\Sigma '}^{11}&=& e^{-2 y} \left(0.0438-0.166 y+0.221 y^2-0.102 y^3+0.0277 y^4\right)\nn \\ 
W_{\Delta }^{00}&=& e^{-2 y} \left(0.126-0.101 y+0.0238 y^2\right)\nn \\ 
W_{\Delta }^{10}=W_{\Delta }^{01}&=& e^{-2 y} \left(0.085-0.068 y+0.0168 y^2\right)\nn \\ 
W_{\Delta }^{11}&=& e^{-2 y} \left(0.0574-0.0459 y+0.0121 y^2\right)\nn \\ 
W_{\Delta \Sigma '}^{00}&=& e^{-2 y} \left(-0.0883+0.186 y-0.104 y^2+0.0164 y^3\right)\nn \\ 
W_{\Delta \Sigma '}^{10}&=& e^{-2 y} \left(-0.0596+0.125 y-0.0717 y^2+0.0114 y^3\right)\nn \\ 
W_{\Delta \Sigma '}^{01}&=& e^{-2 y} \left(-0.0743+0.17 y-0.106 y^2+0.0188 y^3\right)\nn \\ 
W_{\Delta \Sigma '}^{11}&=& e^{-2 y} \left(-0.0501+0.115 y-0.073 y^2+0.0131 y^3\right)\nn 
\end{eqnarray*}

$\mathbf{_{14}^{29}Si}$
\begin{eqnarray*}
W_M^{00}&=& e^{-2 y} \left(33.5-58.5 y+35.1 y^2-8.34 y^3+0.681 y^4\right)\nn \\ 
W_M^{10}=W_M^{01}&=& e^{-2 y} \left(-1.15+2.55 y-2.05 y^2+0.693 y^3-0.0773 y^4\right)\nn \\ 
W_M^{11}&=& e^{-2 y} \left(0.0398-0.106 y+0.108 y^2-0.0498 y^3+0.00876 y^4\right)\nn \\ 
W_{\Sigma ''}^{00}&=& e^{-2 y} \left(0.00273-0.000186 y+0.00447 y^2-0.000152 y^3+0.00183 y^4\right)\nn \\ 
W_{\Sigma ''}^{10}=W_{\Sigma ''}^{01}&=& e^{-2 y} \left(-0.00281+0.00011 y-0.00436 y^2+0.0000821 y^3-0.00169 y^4\right)\nn \\ 
W_{\Sigma ''}^{11}&=& e^{-2 y} \left(0.0029-0.0000299 y+0.00425 y^2-0.0000219 y^3+0.00156 y^4\right)\nn \\ 
W_{\Sigma '}^{00}&=& e^{-2 y} \left(0.00546-0.0216 y+0.0357 y^2-0.0282 y^3+0.00925 y^4\right)\nn \\ 
W_{\Sigma '}^{10}=W_{\Sigma '}^{01}&=& e^{-2 y} \left(-0.00563+0.0224 y-0.0367 y^2+0.0287 y^3-0.00925 y^4\right)\nn \\ 
W_{\Sigma '}^{11}&=& e^{-2 y} \left(0.0058-0.0232 y+0.0378 y^2-0.0293 y^3+0.00924 y^4\right)\nn \\ 
W_{\Delta }^{00}&=& e^{-2 y} \left(0.000434 (2.5- y)^2\right)\nn \\ 
W_{\Delta }^{10}=W_{\Delta }^{01}&=& e^{-2 y} \left(-0.000389 (2.5-y)^2\right)\nn \\ 
W_{\Delta }^{11}&=& e^{-2 y} \left(0.00035 (2.5- y)^2\right)\nn \\ 
W_{\Delta \Sigma '}^{00}&=& e^{-2 y} \left(-0.00385+0.00916 y-0.00806 y^2+0.002 y^3\right)\nn \\ 
W_{\Delta \Sigma '}^{10}&=& e^{-2 y} \left(0.00345-0.00823 y+0.00724 y^2-0.0018 y^3\right)\nn \\ 
W_{\Delta \Sigma '}^{01}&=& e^{-2 y} \left(0.00397-0.00951 y+0.00817 y^2-0.002 y^3\right)\nn \\ 
W_{\Delta \Sigma '}^{11}&=& e^{-2 y} \left(-0.00356+0.00854 y-0.00734 y^2+0.0018 y^3\right)\nn 
\end{eqnarray*}

$\mathbf{_{26}^{56}Fe}$
\begin{eqnarray*}
W_M^{00}&=& e^{-2 y} \left(62.4-160. y+152. y^2-67. y^3+14.4 y^4-1.41 y^5+0.0512 y^6\right)\nn \\ 
W_M^{10}=W_M^{01}&=& e^{-2 y} \left(-4.46+14.6 y-18.1 y^2+10.6 y^3-3.1 y^4+0.421 y^5-0.0204 y^6\right)\nn \\ 
W_M^{11}&=& e^{-2 y} \left(0.318-1.27 y+1.97 y^2-1.49 y^3+0.583 y^4-0.111 y^5+0.00813 y^6\right)\nn 
\end{eqnarray*}

\newpage

\pagebreak

\bibliographystyle{utphys}
\bibliography{midm}

\providecommand{\href}[2]{#2}\begingroup\raggedright\begin{thebibliography}{10}

\bibitem{Han:1997wn}
T.~Han and R.~Hempfling, ``{Messenger sneutrinos as cold dark matter},''
  \href{http://dx.doi.org/10.1016/S0370-2693(97)01205-7}{{\em Phys. Lett. B}
  {\bfseries 415} (1997) 161--169},
  \href{http://arxiv.org/abs/hep-ph/9708264}{{\ttfamily arXiv:hep-ph/9708264}}.

\bibitem{Hall:1997ah}
L.~J. Hall, T.~Moroi, and H.~Murayama, ``{Sneutrino cold dark matter with
  lepton number violation},''
  \href{http://dx.doi.org/10.1016/S0370-2693(98)00196-8}{{\em Phys. Lett.}
  {\bfseries B424} (1998) 305--312},
\href{http://arxiv.org/abs/hep-ph/9712515}{{\ttfamily arXiv:hep-ph/9712515
  [hep-ph]}}.

\bibitem{TuckerSmith:2001hy}
D.~Tucker-Smith and N.~Weiner, ``{Inelastic dark matter},''
  \href{http://dx.doi.org/10.1103/PhysRevD.64.043502}{{\em Phys. Rev.}
  {\bfseries D64} (2001) 043502},
\href{http://arxiv.org/abs/hep-ph/0101138}{{\ttfamily arXiv:hep-ph/0101138
  [hep-ph]}}.

\bibitem{TuckerSmith:2004jv}
D.~Tucker-Smith and N.~Weiner, ``{The Status of inelastic dark matter},''
  \href{http://dx.doi.org/10.1103/PhysRevD.72.063509}{{\em Phys. Rev.}
  {\bfseries D72} (2005) 063509},
\href{http://arxiv.org/abs/hep-ph/0402065}{{\ttfamily arXiv:hep-ph/0402065
  [hep-ph]}}.

\bibitem{Finkbeiner:2007kk}
D.~P. Finkbeiner and N.~Weiner, ``{Exciting Dark Matter and the INTEGRAL/SPI
  511 keV signal},'' \href{http://dx.doi.org/10.1103/PhysRevD.76.083519}{{\em
  Phys. Rev.} {\bfseries D76} (2007) 083519},
\href{http://arxiv.org/abs/astro-ph/0702587}{{\ttfamily arXiv:astro-ph/0702587
  [astro-ph]}}.

\bibitem{Arina:2007tm}
C.~Arina and N.~Fornengo, ``{Sneutrino cold dark matter, a new analysis: Relic
  abundance and detection rates},''
  \href{http://dx.doi.org/10.1088/1126-6708/2007/11/029}{{\em JHEP} {\bfseries
  11} (2007) 029},
\href{http://arxiv.org/abs/0709.4477}{{\ttfamily arXiv:0709.4477 [hep-ph]}}.

\bibitem{Chang:2008gd}
S.~Chang, G.~D. Kribs, D.~Tucker-Smith, and N.~Weiner, ``{Inelastic Dark Matter
  in Light of DAMA/LIBRA},''
  \href{http://dx.doi.org/10.1103/PhysRevD.79.043513}{{\em Phys. Rev.}
  {\bfseries D79} (2009) 043513},
\href{http://arxiv.org/abs/0807.2250}{{\ttfamily arXiv:0807.2250 [hep-ph]}}.

\bibitem{Cui:2009xq}
Y.~Cui, D.~E. Morrissey, D.~Poland, and L.~Randall, ``{Candidates for Inelastic
  Dark Matter},'' \href{http://dx.doi.org/10.1088/1126-6708/2009/05/076}{{\em
  JHEP} {\bfseries 05} (2009) 076},
\href{http://arxiv.org/abs/0901.0557}{{\ttfamily arXiv:0901.0557 [hep-ph]}}.

\bibitem{Fox:2010bu}
P.~J. Fox, G.~D. Kribs, and T.~M.~P. Tait, ``{Interpreting Dark Matter Direct
  Detection Independently of the Local Velocity and Density Distribution},''
  \href{http://dx.doi.org/10.1103/PhysRevD.83.034007}{{\em Phys. Rev.}
  {\bfseries D83} (2011) 034007},
\href{http://arxiv.org/abs/1011.1910}{{\ttfamily arXiv:1011.1910 [hep-ph]}}.

\bibitem{Lin:2010sb}
T.~Lin and D.~P. Finkbeiner, ``{Magnetic Inelastic Dark Matter: Directional
  Signals Without a Directional Detector},''
  \href{http://dx.doi.org/10.1103/PhysRevD.83.083510}{{\em Phys. Rev.}
  {\bfseries D83} (2011) 083510},
\href{http://arxiv.org/abs/1011.3052}{{\ttfamily arXiv:1011.3052
  [astro-ph.CO]}}.

\bibitem{An:2011uq}
H.~An, P.~S.~B. Dev, Y.~Cai, and R.~N. Mohapatra, ``{Sneutrino Dark Matter in
  Gauged Inverse Seesaw Models for Neutrinos},''
  \href{http://dx.doi.org/10.1103/PhysRevLett.108.081806}{{\em Phys. Rev.
  Lett.} {\bfseries 108} (2012) 081806},
\href{http://arxiv.org/abs/1110.1366}{{\ttfamily arXiv:1110.1366 [hep-ph]}}.

\bibitem{Pospelov:2013nea}
M.~Pospelov, N.~Weiner, and I.~Yavin, ``{Dark matter detection in two easy
  steps},'' \href{http://dx.doi.org/10.1103/PhysRevD.89.055008}{{\em Phys.
  Rev.} {\bfseries D89} no.~5, (2014) 055008},
\href{http://arxiv.org/abs/1312.1363}{{\ttfamily arXiv:1312.1363 [hep-ph]}}.

\bibitem{Dienes:2014via}
K.~R. Dienes, J.~Kumar, B.~Thomas, and D.~Yaylali, ``{Dark-Matter Decay as a
  Complementary Probe of Multicomponent Dark Sectors},''
  \href{http://dx.doi.org/10.1103/PhysRevLett.114.051301}{{\em Phys. Rev.
  Lett.} {\bfseries 114} no.~5, (2015) 051301},
\href{http://arxiv.org/abs/1406.4868}{{\ttfamily arXiv:1406.4868 [hep-ph]}}.

\bibitem{Barello:2014uda}
G.~Barello, S.~Chang, and C.~A. Newby, ``{A Model Independent Approach to
  Inelastic Dark Matter Scattering},''
  \href{http://dx.doi.org/10.1103/PhysRevD.90.094027}{{\em Phys. Rev.}
  {\bfseries D90} no.~9, (2014) 094027},
\href{http://arxiv.org/abs/1409.0536}{{\ttfamily arXiv:1409.0536 [hep-ph]}}.

\bibitem{Bramante:2016rdh}
J.~Bramante, P.~J. Fox, G.~D. Kribs, and A.~Martin, ``{Inelastic frontier:
  Discovering dark matter at high recoil energy},''
  \href{http://dx.doi.org/10.1103/PhysRevD.94.115026}{{\em Phys. Rev.}
  {\bfseries D94} no.~11, (2016) 115026},
\href{http://arxiv.org/abs/1608.02662}{{\ttfamily arXiv:1608.02662 [hep-ph]}}.

\bibitem{Krall:2017xij}
R.~Krall and M.~Reece, ``{Last Electroweak WIMP Standing: Pseudo-Dirac Higgsino
  Status and Compact Stars as Future Probes},''
  \href{http://dx.doi.org/10.1088/1674-1137/42/4/043105}{{\em Chin. Phys.}
  {\bfseries C42} no.~4, (2018) 043105},
\href{http://arxiv.org/abs/1705.04843}{{\ttfamily arXiv:1705.04843 [hep-ph]}}.

\bibitem{Song:2021yar}
N.~Song, S.~Nagorny, and A.~C. Vincent, ``{Pushing the frontier of WIMPy
  inelastic dark matter: Journey to the end of the periodic table},''
  \href{http://dx.doi.org/10.1103/PhysRevD.104.103032}{{\em Phys. Rev. D}
  {\bfseries 104} no.~10, (2021) 103032},
  \href{http://arxiv.org/abs/2104.09517}{{\ttfamily arXiv:2104.09517
  [hep-ph]}}.

\bibitem{Bell:2018pkk}
N.~F. Bell, G.~Busoni, and S.~Robles, ``{Heating up Neutron Stars with
  Inelastic Dark Matter},''
  \href{http://dx.doi.org/10.1088/1475-7516/2018/09/018}{{\em JCAP} {\bfseries
  09} (2018) 018}, \href{http://arxiv.org/abs/1807.02840}{{\ttfamily
  arXiv:1807.02840 [hep-ph]}}.

\bibitem{Catena:2018vzc}
R.~Catena and F.~Hellstr\"om, ``{New constraints on inelastic dark matter from
  IceCube},'' \href{http://dx.doi.org/10.1088/1475-7516/2018/10/039}{{\em JCAP}
  {\bfseries 10} (2018) 039}, \href{http://arxiv.org/abs/1808.08082}{{\ttfamily
  arXiv:1808.08082 [astro-ph.CO]}}.

\bibitem{Pospelov:2019vuf}
M.~Pospelov, S.~Rajendran, and H.~Ramani, ``{Metastable Nuclear Isomers as Dark
  Matter Accelerators},''
  \href{http://dx.doi.org/10.1103/PhysRevD.101.055001}{{\em Phys. Rev. D}
  {\bfseries 101} no.~5, (2020) 055001},
  \href{http://arxiv.org/abs/1907.00011}{{\ttfamily arXiv:1907.00011
  [hep-ph]}}.

\bibitem{Lehnert:2019tuw}
B.~Lehnert, H.~Ramani, M.~Hult, G.~Lutter, M.~Pospelov, S.~Rajendran, and
  K.~Zuber, ``{Search for Dark Matter Induced Deexcitation of $^{180}$Ta$\rm
  ^m$},'' \href{http://dx.doi.org/10.1103/PhysRevLett.124.181802}{{\em Phys.
  Rev. Lett.} {\bfseries 124} no.~18, (2020) 181802},
  \href{http://arxiv.org/abs/1911.07865}{{\ttfamily arXiv:1911.07865
  [astro-ph.CO]}}.

\bibitem{Broerman:2020hfj}
B.~Broerman, M.~Laubenstein, S.~Nagorny, N.~Song, and A.~C. Vincent, ``{A
  search for rare and induced nuclear decays in hafnium},''
  \href{http://dx.doi.org/10.1016/j.nuclphysa.2021.122212}{{\em Nucl. Phys. A}
  {\bfseries 1012} (2021) 122212},
  \href{http://arxiv.org/abs/2012.08339}{{\ttfamily arXiv:2012.08339
  [nucl-ex]}}.

\bibitem{Tsai:2019buq}
Y.-D. Tsai, P.~deNiverville, and M.~X. Liu, ``{Dark Photon and Muon $g-2$
  Inspired Inelastic Dark Matter Models at the High-Energy Intensity
  Frontier},'' \href{http://dx.doi.org/10.1103/PhysRevLett.126.181801}{{\em
  Phys. Rev. Lett.} {\bfseries 126} no.~18, (2021) 181801},
  \href{http://arxiv.org/abs/1908.07525}{{\ttfamily arXiv:1908.07525
  [hep-ph]}}.

\bibitem{Bell:2021zkr}
N.~F. Bell, J.~B. Dent, B.~Dutta, S.~Ghosh, J.~Kumar, and J.~L. Newstead,
  ``{Low-mass inelastic dark matter direct detection via the Migdal effect},''
  \href{http://dx.doi.org/10.1103/PhysRevD.104.076013}{{\em Phys. Rev. D}
  {\bfseries 104} no.~7, (2021) 076013},
  \href{http://arxiv.org/abs/2103.05890}{{\ttfamily arXiv:2103.05890
  [hep-ph]}}.

\bibitem{Bell:2021xff}
N.~F. Bell, J.~B. Dent, B.~Dutta, S.~Ghosh, J.~Kumar, J.~L. Newstead, and I.~M.
  Shoemaker, ``{Cosmic-ray upscattered inelastic dark matter},''
  \href{http://dx.doi.org/10.1103/PhysRevD.104.076020}{{\em Phys. Rev. D}
  {\bfseries 104} (2021) 076020},
  \href{http://arxiv.org/abs/2108.00583}{{\ttfamily arXiv:2108.00583
  [hep-ph]}}.

\bibitem{CarrilloGonzalez:2021lxm}
M.~Carrillo~Gonz\'alez and N.~Toro, ``{Cosmology and signals of light
  pseudo-Dirac dark matter},''
  \href{http://dx.doi.org/10.1007/JHEP04(2022)060}{{\em JHEP} {\bfseries 04}
  (2022) 060}, \href{http://arxiv.org/abs/2108.13422}{{\ttfamily
  arXiv:2108.13422 [hep-ph]}}.

\bibitem{Filimonova:2022pkj}
A.~Filimonova, S.~Junius, L.~Lopez~Honorez, and S.~Westhoff, ``{Inelastic Dirac
  dark matter},'' \href{http://dx.doi.org/10.1007/JHEP06(2022)048}{{\em JHEP}
  {\bfseries 06} (2022) 048}, \href{http://arxiv.org/abs/2201.08409}{{\ttfamily
  arXiv:2201.08409 [hep-ph]}}.

\bibitem{Bell:2022yxn}
N.~F. Bell, J.~B. Dent, B.~Dutta, J.~Kumar, and J.~L. Newstead, ``{Low-Mass
  dark matter (in)direct detection with inelastic scattering},''
  \href{http://arxiv.org/abs/2208.08020}{{\ttfamily arXiv:2208.08020
  [hep-ph]}}.

\bibitem{Berlin:2023qco}
A.~Berlin, G.~Krnjaic, and E.~Pinetti, ``{Reviving MeV-GeV Indirect Detection
  with Inelastic Dark Matter},''
  \href{http://arxiv.org/abs/2311.00032}{{\ttfamily arXiv:2311.00032
  [hep-ph]}}.

\bibitem{Bramante:2023ddr}
J.~Bramante, M.~Diamond, and J.~L. Kim, ``{The Effect of Multiple Cooling
  Channels on the Formation of Dark Compact Objects},''
  \href{http://arxiv.org/abs/2309.13148}{{\ttfamily arXiv:2309.13148
  [hep-ph]}}.

\bibitem{Dienes:2023uve}
K.~R. Dienes, J.~L. Feng, M.~Fieg, F.~Huang, S.~J. Lee, and B.~Thomas,
  ``{Extending the discovery potential for inelastic-dipole dark matter with
  FASER},'' \href{http://dx.doi.org/10.1103/PhysRevD.107.115006}{{\em Phys.
  Rev. D} {\bfseries 107} no.~11, (2023) 115006},
  \href{http://arxiv.org/abs/2301.05252}{{\ttfamily arXiv:2301.05252
  [hep-ph]}}.

\bibitem{Asai:2023dzs}
K.~Asai, S.~Iwamoto, M.~Perelstein, Y.~Sakaki, and D.~Ueda, ``{Sub-GeV dark
  matter search at ILC beam dumps},''
  \href{http://arxiv.org/abs/2301.03816}{{\ttfamily arXiv:2301.03816
  [hep-ph]}}.

\bibitem{Jodlowski:2023ohn}
K.~Jodlowski, ``{Looking forward to inelastic DM with electromagnetic form
  factors at FASER and beam dump experiments},''
  \href{http://arxiv.org/abs/2305.16781}{{\ttfamily arXiv:2305.16781
  [hep-ph]}}.

\bibitem{Lu:2023cet}
C.-T. Lu, J.~Tu, and L.~Wu, ``{Probing Inelastic Dark Matter at the LHC, FASER
  and STCF},'' \href{http://arxiv.org/abs/2309.00271}{{\ttfamily
  arXiv:2309.00271 [hep-ph]}}.

\bibitem{Chauhan:2023zuf}
B.~Chauhan, M.~H. Reno, C.~Rott, and I.~Sarcevic, ``{Neutrino constraints on
  inelastic dark matter captured in the Sun},''
  \href{http://arxiv.org/abs/2308.16134}{{\ttfamily arXiv:2308.16134
  [hep-ph]}}.

\bibitem{Heeba:2023bik}
S.~Heeba, T.~Lin, and K.~Schutz, ``{Inelastic freeze-in},''
  \href{http://dx.doi.org/10.1103/PhysRevD.108.095016}{{\em Phys. Rev. D}
  {\bfseries 108} no.~9, (2023) 095016},
  \href{http://arxiv.org/abs/2304.06072}{{\ttfamily arXiv:2304.06072
  [hep-ph]}}.

\bibitem{Bell:2022dbf}
N.~F. Bell, J.~B. Dent, B.~Dutta, J.~Kumar, and J.~L. Newstead, ``{Indirect
  detection of low mass dark matter in direct detection experiments with
  inelastic scattering},''
  \href{http://dx.doi.org/10.1103/PhysRevD.106.103016}{{\em Phys. Rev. D}
  {\bfseries 106} no.~10, (2022) 103016},
  \href{http://arxiv.org/abs/2208.08020}{{\ttfamily arXiv:2208.08020
  [hep-ph]}}.

\bibitem{PandaX:2022djq}
{\bfseries PandaX} Collaboration, Y.~Yuan {\em et~al.}, ``{A search for
  two-component Majorana dark matter in a simplified model using the full
  exposure data of PandaX-II experiment},''
  \href{http://dx.doi.org/10.1016/j.physletb.2022.137254}{{\em Phys. Lett. B}
  {\bfseries 832} (2022) 137254},
  \href{http://arxiv.org/abs/2205.08066}{{\ttfamily arXiv:2205.08066
  [hep-ex]}}.

\bibitem{XENON:2022avm}
{\bfseries XENON} Collaboration, E.~Aprile {\em et~al.}, ``{Effective Field
  Theory and Inelastic Dark Matter Results from XENON1T},''
  \href{http://arxiv.org/abs/2210.07591}{{\ttfamily arXiv:2210.07591
  [hep-ex]}}.

\bibitem{Gu:2022vgb}
Y.~Gu, L.~Wu, and B.~Zhu, ``{Detection of inelastic dark matter via electron
  recoils in SENSEI},''
  \href{http://dx.doi.org/10.1103/PhysRevD.106.075004}{{\em Phys. Rev. D}
  {\bfseries 106} no.~7, (2022) 075004},
  \href{http://arxiv.org/abs/2203.06664}{{\ttfamily arXiv:2203.06664
  [hep-ph]}}.

\bibitem{Bell:2020bes}
N.~F. Bell, J.~B. Dent, B.~Dutta, S.~Ghosh, J.~Kumar, and J.~L. Newstead,
  ``{Explaining the XENON1T excess with Luminous Dark Matter},''
  \href{http://dx.doi.org/10.1103/PhysRevLett.125.161803}{{\em Phys. Rev.
  Lett.} {\bfseries 125} no.~16, (2020) 161803},
  \href{http://arxiv.org/abs/2006.12461}{{\ttfamily arXiv:2006.12461
  [hep-ph]}}.

\bibitem{Bramante:2020zos}
J.~Bramante and N.~Song, ``{Electric But Not Eclectic: Thermal Relic Dark
  Matter for the XENON1T Excess},''
  \href{http://dx.doi.org/10.1103/PhysRevLett.125.161805}{{\em Phys. Rev.
  Lett.} {\bfseries 125} no.~16, (2020) 161805},
  \href{http://arxiv.org/abs/2006.14089}{{\ttfamily arXiv:2006.14089
  [hep-ph]}}.

\bibitem{Baryakhtar:2020rwy}
M.~Baryakhtar, A.~Berlin, H.~Liu, and N.~Weiner, ``{Electromagnetic signals of
  inelastic dark matter scattering},''
  \href{http://dx.doi.org/10.1007/JHEP06(2022)047}{{\em JHEP} {\bfseries 06}
  (2022) 047}, \href{http://arxiv.org/abs/2006.13918}{{\ttfamily
  arXiv:2006.13918 [hep-ph]}}.

\bibitem{Bloch:2020uzh}
I.~M. Bloch, A.~Caputo, R.~Essig, D.~Redigolo, M.~Sholapurkar, and T.~Volansky,
  ``{Exploring new physics with O(keV) electron recoils in direct detection
  experiments},'' \href{http://dx.doi.org/10.1007/JHEP01(2021)178}{{\em JHEP}
  {\bfseries 01} (2021) 178}, \href{http://arxiv.org/abs/2006.14521}{{\ttfamily
  arXiv:2006.14521 [hep-ph]}}.

\bibitem{Emken:2021vmf}
T.~Emken, J.~Frerick, S.~Heeba, and F.~Kahlhoefer, ``{Electron recoils from
  terrestrial upscattering of inelastic dark matter},''
  \href{http://dx.doi.org/10.1103/PhysRevD.105.055023}{{\em Phys. Rev. D}
  {\bfseries 105} no.~5, (2022) 055023},
  \href{http://arxiv.org/abs/2112.06930}{{\ttfamily arXiv:2112.06930
  [hep-ph]}}.

\bibitem{XENON:2020rca}
{\bfseries XENON} Collaboration, E.~Aprile {\em et~al.}, ``{Excess electronic
  recoil events in XENON1T},''
  \href{http://dx.doi.org/10.1103/PhysRevD.102.072004}{{\em Phys. Rev. D}
  {\bfseries 102} no.~7, (2020) 072004},
  \href{http://arxiv.org/abs/2006.09721}{{\ttfamily arXiv:2006.09721
  [hep-ex]}}.

\bibitem{Feldstein:2010su}
B.~Feldstein, P.~W. Graham, and S.~Rajendran, ``{Luminous Dark Matter},''
  \href{http://dx.doi.org/10.1103/PhysRevD.82.075019}{{\em Phys. Rev.}
  {\bfseries D82} (2010) 075019},
\href{http://arxiv.org/abs/1008.1988}{{\ttfamily arXiv:1008.1988 [hep-ph]}}.

\bibitem{Eby:2019mgs}
J.~Eby, P.~J. Fox, R.~Harnik, and G.~D. Kribs, ``{Luminous Signals of Inelastic
  Dark Matter in Large Detectors},''
  \href{http://dx.doi.org/10.1007/JHEP09(2019)115}{{\em JHEP} {\bfseries 09}
  (2019) 115},
\href{http://arxiv.org/abs/1904.09994}{{\ttfamily arXiv:1904.09994 [hep-ph]}}.

\bibitem{Fox:2014moa}
P.~J. Fox, G.~D. Kribs, and A.~Martin, ``{Split Dirac Supersymmetry: An
  Ultraviolet Completion of Higgsino Dark Matter},''
  \href{http://dx.doi.org/10.1103/PhysRevD.90.075006}{{\em Phys. Rev.}
  {\bfseries D90} no.~7, (2014) 075006},
\href{http://arxiv.org/abs/1405.3692}{{\ttfamily arXiv:1405.3692 [hep-ph]}}.

\bibitem{Plestid:2020vqf}
R.~Plestid, ``{Luminous solar neutrinos I: Dipole portals},''
  \href{http://dx.doi.org/10.1103/PhysRevD.104.075027}{{\em Phys. Rev. D}
  {\bfseries 104} (2021) 075027},
  \href{http://arxiv.org/abs/2010.04193}{{\ttfamily arXiv:2010.04193
  [hep-ph]}}.

\bibitem{Plestid:2020ssy}
R.~Plestid, ``{Luminous solar neutrinos II: Mass-mixing portals},''
  \href{http://dx.doi.org/10.1103/PhysRevD.104.075028}{{\em Phys. Rev. D}
  {\bfseries 104} (2021) 075028},
  \href{http://arxiv.org/abs/2010.09523}{{\ttfamily arXiv:2010.09523
  [hep-ph]}}. [Erratum: Phys.Rev.D 105, 099901 (2022)].

\bibitem{Kopp:2009qt}
J.~Kopp, T.~Schwetz, and J.~Zupan, ``{Global interpretation of direct Dark
  Matter searches after CDMS-II results},''
  \href{http://dx.doi.org/10.1088/1475-7516/2010/02/014}{{\em JCAP} {\bfseries
  02} (2010) 014}, \href{http://arxiv.org/abs/0912.4264}{{\ttfamily
  arXiv:0912.4264 [hep-ph]}}.

\bibitem{Chang:2010en}
S.~Chang, N.~Weiner, and I.~Yavin, ``{Magnetic Inelastic Dark Matter},''
  \href{http://dx.doi.org/10.1103/PhysRevD.82.125011}{{\em Phys. Rev.}
  {\bfseries D82} (2010) 125011},
\href{http://arxiv.org/abs/1007.4200}{{\ttfamily arXiv:1007.4200 [hep-ph]}}.

\bibitem{Vahsen:2020pzb}
S.~E. Vahsen {\em et~al.}, ``{CYGNUS: Feasibility of a nuclear recoil
  observatory with directional sensitivity to dark matter and neutrinos},''
  \href{http://arxiv.org/abs/2008.12587}{{\ttfamily arXiv:2008.12587
  [physics.ins-det]}}.

\bibitem{Asadi:2023toappear}
P.~Asadi, G.~D. Kribs, and C.~Mantel, ``to appear,''.

\bibitem{LEPchargebound1}
{\bfseries ALEPH, DELPHI, L3, and OPAL} Collaboration, ``{Combined LEP Chargino
  Results, up to 208 GeV for large $m0$}.''
  \href{https://lepsusy.web.cern.ch/lepsusy/www/inos_moriond01/charginos_pub.html}{LEPSUSYWG/01-03.1}.

\bibitem{LEPchargebound2}
{\bfseries ALEPH, DELPHI, L3, and OPAL} Collaboration, ``{Combined LEP Chargino
  Results, up to 208 GeV for low DM}.''
  \href{https://lepsusy.web.cern.ch/lepsusy/www/inoslowdmsummer02/charginolowdm_pub.html}{
  LEPSUSYWG/02-04.1 }.

\bibitem{DelNobile:2021wmp}
E.~Del~Nobile, ``{The Theory of Direct Dark Matter Detection: A Guide to
  Computations},'' \href{http://dx.doi.org/10.1007/978-3-030-95228-0}{{\em
  Lect. Notes Phys.} {\bfseries 996} (5, 2022) },
  \href{http://arxiv.org/abs/2104.12785}{{\ttfamily arXiv:2104.12785
  [hep-ph]}}.

\bibitem{Fitzpatrick:2012ix}
A.~L. Fitzpatrick, W.~Haxton, E.~Katz, N.~Lubbers, and Y.~Xu, ``{The Effective
  Field Theory of Dark Matter Direct Detection},''
  \href{http://dx.doi.org/10.1088/1475-7516/2013/02/004}{{\em JCAP} {\bfseries
  1302} (2013) 004},
\href{http://arxiv.org/abs/1203.3542}{{\ttfamily arXiv:1203.3542 [hep-ph]}}.

\bibitem{Anand:2013yka}
N.~Anand, A.~L. Fitzpatrick, and W.~C. Haxton, ``{Weakly interacting massive
  particle-nucleus elastic scattering response},''
  \href{http://dx.doi.org/10.1103/PhysRevC.89.065501}{{\em Phys. Rev.}
  {\bfseries C89} no.~6, (2014) 065501},
\href{http://arxiv.org/abs/1308.6288}{{\ttfamily arXiv:1308.6288 [hep-ph]}}.

\bibitem{nuclearwallet}
{National Nuclear Data Center}, ``Nuclear wallet card.''.

\bibitem{STONE200575}
N.~Stone, ``Table of nuclear magnetic dipole and electric quadrupole moments,''
  \href{http://dx.doi.org/https://doi.org/10.1016/j.adt.2005.04.001}{{\em
  Atomic Data and Nuclear Data Tables} {\bfseries 90} no.~1, (2005) 75--176}.
  \url{https://www.sciencedirect.com/science/article/pii/S0092640X05000239}.

\bibitem{core}
W.~McDonough,
  \href{http://dx.doi.org/https://doi.org/10.1016/B978-0-08-095975-7.00201-1}{``3.16
  - compositional model for the earth's core,''} in {\em Treatise on
  Geochemistry (Second Edition)}, H.~D. Holland and K.~K. Turekian, eds.,
  pp.~559 -- 597.
\newblock Elsevier, Oxford, second edition~ed., 2014.
\newblock
  \url{https://www.sciencedirect.com/science/article/pii/B9780080959757002151}.

\bibitem{mantle}
H.~Palme and H.~O'Neill,
  \href{http://dx.doi.org/https://doi.org/10.1016/B978-0-08-095975-7.00201-1}{``3.1
  - cosmochemical estimates of mantle composition,''} in {\em Treatise on
  Geochemistry (Second Edition)}, H.~D. Holland and K.~K. Turekian, eds., pp.~1
  -- 39.
\newblock Elsevier, Oxford, second edition~ed., 2014.
\newblock
  \url{http://www.sciencedirect.com/science/article/pii/B9780080959757002011}.

\bibitem{crust}
R.~Rudnick and S.~Gao,
  \href{http://dx.doi.org/https://doi.org/10.1016/B978-0-08-095975-7.00301-6}{``4.1
  - composition of the continental crust,''} in {\em Treatise on Geochemistry
  (Second Edition)}, H.~D. Holland and K.~K. Turekian, eds., pp.~1 -- 51.
\newblock Elsevier, Oxford, second edition~ed., 2014.
\newblock
  \url{http://www.sciencedirect.com/science/article/pii/B9780080959757003016}.

\bibitem{Eby_IsotopeResponses}
J.~Eby, P.~J. Fox, and G.~D. Kribs, ``{IsotopeResponses}.''
\newblock \url{https://github.com/joshaeby/IsotopeResponses}.

\bibitem{CYGNUSww}
{\bfseries CYGNUS} Collaboration, ``{CYGNUS -- A world wide directional dark
  matter detection experiment}.'' \url{https://www.phys.hawaii.edu/cygnus/}.

\bibitem{SABRETDR}
{\bfseries SABRE South} Collaboration, ``{The SABRE South Technical Design
  Report}.''
  \url{https://darkmatteraustralia.atlassian.net/wiki/spaces/SABREPUBLIC/pages/973209623/Publications}.

\bibitem{SABRE:2018lfp}
{\bfseries SABRE} Collaboration, M.~Antonello {\em et~al.}, ``{The SABRE
  project and the SABRE Proof-of-Principle},''
  \href{http://dx.doi.org/10.1140/epjc/s10052-019-6860-y}{{\em Eur. Phys. J. C}
  {\bfseries 79} no.~4, (2019) 363},
  \href{http://arxiv.org/abs/1806.09340}{{\ttfamily arXiv:1806.09340
  [physics.ins-det]}}.

\bibitem{LZ:2022lsv}
{\bfseries LZ} Collaboration, J.~Aalbers {\em et~al.}, ``{First Dark Matter
  Search Results from the LUX-ZEPLIN (LZ) Experiment},''
  \href{http://dx.doi.org/10.1103/PhysRevLett.131.041002}{{\em Phys. Rev.
  Lett.} {\bfseries 131} no.~4, (2023) 041002},
  \href{http://arxiv.org/abs/2207.03764}{{\ttfamily arXiv:2207.03764
  [hep-ex]}}.

\bibitem{Freese:2012xd}
K.~Freese, M.~Lisanti, and C.~Savage, ``{Colloquium: Annual modulation of dark
  matter},'' \href{http://dx.doi.org/10.1103/RevModPhys.85.1561}{{\em Rev. Mod.
  Phys.} {\bfseries 85} (2013) 1561--1581},
  \href{http://arxiv.org/abs/1209.3339}{{\ttfamily arXiv:1209.3339
  [astro-ph.CO]}}.

\bibitem{Vahsen:2021gnb}
S.~E. Vahsen, C.~A.~J. O'Hare, and D.~Loomba, ``{Directional Recoil
  Detection},''
  \href{http://dx.doi.org/10.1146/annurev-nucl-020821-035016}{{\em Ann. Rev.
  Nucl. Part. Sci.} {\bfseries 71} (2021) 189--224},
  \href{http://arxiv.org/abs/2102.04596}{{\ttfamily arXiv:2102.04596
  [physics.ins-det]}}.

\bibitem{Catena:2015uha}
R.~Catena and B.~Schwabe, ``{Form factors for dark matter capture by the Sun in
  effective theories},''
  \href{http://dx.doi.org/10.1088/1475-7516/2015/04/042}{{\em JCAP} {\bfseries
  1504} no.~04, (2015) 042},
\href{http://arxiv.org/abs/1501.03729}{{\ttfamily arXiv:1501.03729 [hep-ph]}}.

\bibitem{Eby:2024toappear}
J.~Eby, P.~J. Fox, and G.~D. Kribs, ``to appear,''.

\bibitem{Dreiner:2008tw}
H.~K. Dreiner, H.~E. Haber, and S.~P. Martin, ``{Two-component spinor
  techniques and Feynman rules for quantum field theory and supersymmetry},''
  \href{http://dx.doi.org/10.1016/j.physrep.2010.05.002}{{\em Phys. Rept.}
  {\bfseries 494} (2010) 1--196},
  \href{http://arxiv.org/abs/0812.1594}{{\ttfamily arXiv:0812.1594 [hep-ph]}}.

\bibitem{Donnelly:1979ezn}
T.~W. Donnelly and W.~C. Haxton, ``{Multipole operators in semileptonic weak
  and electromagnetic interactions with nuclei},''
\href{http://dx.doi.org/10.1016/0092-640X(79)90003-2}{{\em Atom. Data Nucl.
  Data Tabl.} {\bfseries 23} (1979) 103--176}.

\bibitem{Johnson:2018hrx}
C.~W. Johnson, W.~E. Ormand, K.~S. McElvain, and H.~Shan, ``{BIGSTICK: A
  flexible configuration-interaction shell-model code},''
  \href{http://arxiv.org/abs/1801.08432}{{\ttfamily arXiv:1801.08432
  [physics.comp-ph]}}.

\bibitem{Oxbash}
W.~R. B.~Brown, A.~Etchegoyen, ``Computer code oxbash: the oxford
  university-buenos aires-msu shell model code,'' Tech. Rep. 524, Michigan
  State University Cyclotron Laboratory Report, 1985.

\bibitem{NuShell}
B.~A. Brown and W.~D.~M. Rae, ``{The Shell-Model Code NuShellX@MSU},''
  \href{http://dx.doi.org/10.1016/j.nds.2014.07.022}{{\em Nucl. Data Sheets}
  {\bfseries 120} (2014) 115--118}.

\end{thebibliography}\endgroup

\newpage

\end{document}